\documentclass{nature}
\usepackage{graphicx}
\usepackage{caption}
\usepackage{hyperref}
\def \aap {Astron. Astrophys.} % Astron. Astrophys.
\def \aj {Astron. J.} % Astron. J.

\def \apj {Astrophys. J.} % Astrophys. J.
\def \apjl {Astrophys. J. Lett.} % Astrophys. J. Lett.
\def \i3etm {IEEE Trans. Magn.}
\def \i3etns {IEEE Trans. Nucl. Sci.}
\def \i3etps {IEEE Trans. Plasma Sci.}

\def \jjap1 {Jpn. J. Appl. Phys., Part 1}
\def \jjap2 {Jpn. J. Appl. Phys., Part 2}

\def \mnras {Mon. Not. R. Astron. Soc.} % Mon. Not. R. Astron. Soc.

\def \nat {Nature} % Nature

\def \pasp {Publ. Astron. Soc. Pac.} % Publ. Astron. Soc. Pac.

\def \usnos4 {U.S. Nav. Obs., Ser. 4}

\title{A map of the large day-night temperature gradient of a super-Earth exoplanet}

\author{Brice-Olivier Demory$^1$, Michael Gillon$^2$, Julien de Wit$^3$, Nikku Madhusudhan$^4$, Emeline Bolmont$^5$, Kevin Heng$^6$, Tiffany Kataria$^7$, Nikole Lewis$^8$,  Renyu Hu$^{9,10}$, Jessica Krick$^{11}$, Vlada Stamenkovi\'c$^{9,10}$, Bj\"orn Benneke$^{10}$, Stephen Kane$^{12}$ \& Didier Queloz$^1$}

\begin{document}

\providecommand{\bibinfo}[2]{#2}
\captionsetup[figure]{labelfont=bf,labelsep=none}
\captionsetup[table]{labelfont=bf,labelsep=none}

\maketitle

\begin{affiliations}
\footnotesize
\item Astrophysics Group, Cavendish Laboratory, J.J. Thomson Avenue, Cambridge CB3 0HE, UK.
\item Institut d'Astrophysique et de G\'eophysique, Universit\'e of Li\`ege, all\'ee du 6 Aout 17, 4000 Li\`ege, Belgium.
\item Department of Earth, Atmospheric and Planetary Sciences, Massachusetts Institute of Technology, 77 Massachusetts Avenue, Cambridge, MA 02139, USA.
\item	Institute of Astronomy, University of Cambridge, Cambridge CB3 0HA, UK.
\item NaXys, Department of Mathematics, University of Namur, 8 Rempart de la Vierge, 5000 Namur, Belgium.
\item University of Bern, Center for Space and Habitability, Sidlerstrasse 5, CH-3012, Bern, Switzerland.
\item Astrophysics Group, School of Physics, University of Exeter, Stocker Road, Exeter EX4 4QL, UK.
\item Space Telescope Science Institute, Baltimore, MD 21218, USA.
\item Jet Propulsion Laboratory, California Institute of Technology, Pasadena, CA 91109, USA.
\item Division of Geological and Planetary Sciences, California Institute of Technology, Pasadena, CA 91125, USA.
\item Spitzer Science Center, MS 220-6, California Institute of Technology, Jet Propulsion Laboratory, Pasadena, CA 91125, USA.
\item Department of Physics \& Astronomy, San Francisco State University, 1600 Holloway Avenue, San Francisco, CA 94132, USA.

\end{affiliations}

\begin{abstract}
Over the past decade, observations of giant exoplanets (Jupiter-size) have provided key insights into their atmospheres\cite{Burrows:2014a,Heng:2015a}, but the properties of lower-mass exoplanets (sub-Neptune) remain largely unconstrained because of the challenges of observing small planets. Numerous efforts to observe the spectra of super-Earths\---exoplanets with masses of one to ten times that of Earth\---have so far revealed only featureless spectra\cite{Knutson:2014a}. Here we report a longitudinal thermal brightness map of the nearby transiting super-Earth 55 Cancri e\cite{Demory:2011,Winn:2011a} revealing highly asymmetric dayside thermal emission and a strong day-night temperature contrast. Dedicated space-based monitoring of the planet in the infrared revealed a modulation of the thermal flux as 55 Cancri e revolves around its star in a tidally locked configuration. These observations reveal a hot spot that is located 41 $\pm$ 12 degrees east of the substellar point (the point at which incident light from the star is perpendicular to the surface of the planet). From the orbital phase curve, we also constrain the nightside brightness temperature of the planet to 1380 $\pm$ 400 kelvin and the temperature of the warmest hemisphere (centred on the hot spot) to be about 1300 kelvin hotter (2700 $\pm$ 270 kelvin) at a wavelength of 4.5 microns, which indicates inefficient heat redistribution from the dayside to the nightside. Our observations are consistent with either an optically thick atmosphere with heat recirculation confined to the planetary dayside, or a planet devoid of atmosphere with low-viscosity magma flows at the surface\cite{Solomatov:2007}.
\end{abstract}

We observed the super-Earth 55\,Cancri\,e for 75 hours in total between Jun 15 and Jul 15 2013 in the 4.5-$\mu$m channel of the {\it Spitzer Space Telescope} Infrared Array Camera (IRAC). The observations were split in 8 continuous visits, each of them spanning 9-hours and corresponding to half of 55\,Cancri\,e's 18-hour orbital period. We acquired a total of 4,981,760 frames in subarray-mode with an individual 0.02-s integration time. We extract the photometric time-series from the raw frames using an aperture photometry pipeline previously described in the literature\cite{Demory:2011}. Each of the 8 resulting light curves exhibit periodic flux variations due to the strong intra-pixel sensitivity of the IRAC detector combined to {\it Spitzer}'s pointing wobble. The data reduction of this dataset has already been published elsewhere\cite{Demory:2015a} but a summary can be found in the Methods section.

We analyse the light curves with a Markov Chain Monte Carlo (MCMC) algorithm presented in a past study\cite{Gillon:2012a}. We simultaneously fit the 8 half phase-curves and a model of the detector systematics. Our MCMC algorithm includes an implementation of a pixel-level correction\cite{Stevenson:2012a} and propagates the contribution from correlated noise in the data to the system best-fit parameters. In our implementation of the method, we build a sub-pixel mesh made of $n^2$ grid points, evenly distributed along the $x$ and $y$ axes. Similar to a previous study\cite{Lanotte:2014}, we find that the point response function's (PRF) full width at half maximum (FWHM) along the $x$ and $y$ axes of the detector evolve with time and allows further improvement on the systematics correction. We thus combine the pixel-mapping algorithm to a linear function of the PRF's FWHM along each axis. We find this model to provide our best correction to the data. The free parameters in the MCMC fit are the phase-curve amplitude and offset, which is the angle between the peak of the modulation and the substellar point, the occultation depth, the planetary impact parameter, orbital period, transit centre and transit depth. The phase-curve functional form used in this fit is detailed in the Methods section. We combine the datapoints per 30-s bins for computing efficiency, which has been previously shown to have no incidence on the derived parameters\cite{Deming:2015,Demory:2015a}. We find an average photometric precision of 363 parts-per-million (ppm) per 30s and evaluate the level of correlated noise in the data for each dataset following a time-averaging technique\cite{Pont:2006b}. Results from the MCMC fits are shown on Table~\ref{tab:res}. We perform two additional analyses of this dataset (see Methods) using a different model for the pixel-level correction, which results in phase-curve parameters in agreement with our main analysis.

\begin{figure}
\centering
\includegraphics[width=\textwidth]{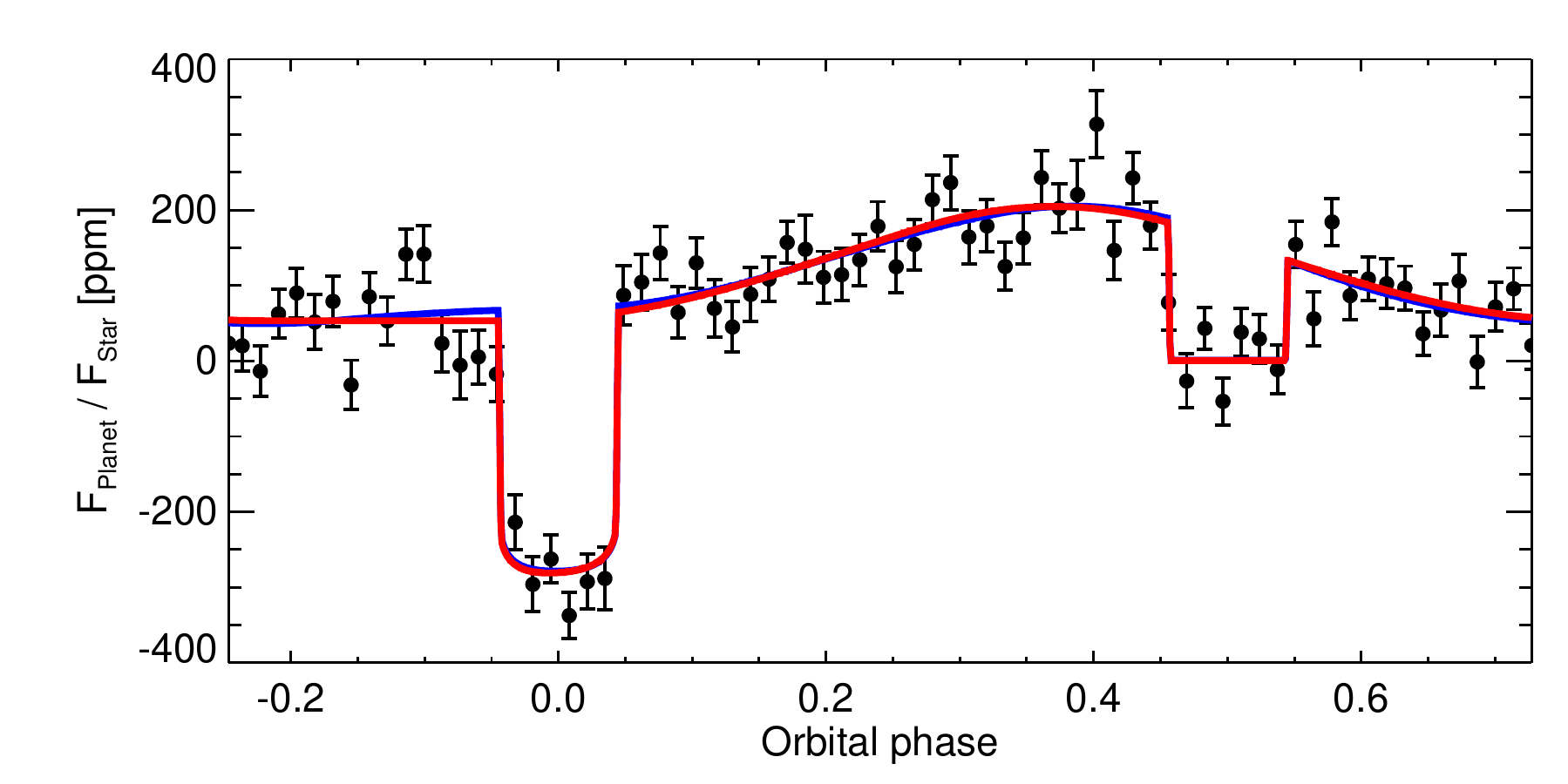}
\caption{\label{fig:lc} | {\bf 55\,Cancri\,e Spitzer/IRAC 4.5$\mu$m phase curve.} Photometry for all 8 datasets combined and folded on 55\,Cancri\,e's 0.74-day orbital period. Black filled circles are data binned per 15 minutes. The best-fit model using a three-longitudinal-band model is shown in red while the best-fit model using a one-longitudinal-band model is shown in blue. The error bars are the standard deviation of the mean within each orbital phase bin.}
\end{figure}

The combined light curve (Figure~\ref{fig:lc}) exhibits a flux increase starting slightly before the transit and reaching a maximum at 2.1$\pm$0.6 hours before opposition. We find a phase-curve peak amplitude of 197$\pm$34 ppm, a minimum of 48$\pm$34 ppm and an occultation depth of 154$\pm$23 ppm (mid-eclipse).

We find that stellar variability could not cause the observed phase variation. The host is known as an old, quiet star with a rotation period of 42 days showing on rare occasions variability at the 6 milli-magnitude level, corresponding to a $<$1\% coverage in star spots\cite{Fischer:2008}. The periodic modulation that we observe is equal to the planetary orbital period and shows a shape that remains consistent over the 4 weeks of the {\it Spitzer} observations. At infrared wavelengths, the effect of starspots on the photometry is dramatically reduced\cite{Berta:2011} but it is still possible that a 1\% spot coverage could produce a signal of the order of $\sim$200 ppm. However the periodicity of the signal produced by such a starspot would be similar to the stellar rotation. 

We also investigate the incidence of the ellipsoidal effects\cite{Mazeh:2010a} caused by 55\,Cancri\,e on its host star and find an expected amplitude of 0.6 ppm. The reciprocal effect from the host star on the planet would translate to an effect of $\sim$1 ppm\cite{Budaj:2011}. None of these features would be detectable in our dataset. In addition, ellipsoidal variations have a frequency that is twice the one of the planet's orbital period. For these two reasons, we discard the possibility that ellipsoidal variations are at the origin of the observed signal. 

An alternate possibility to mimic the orbital phase-curve would be a scenario where 55\,Cancri\,e induces starspots on the stellar surface through magnetic field interactions, which would produce a photometric modulation synchronised with the planet's orbital period\cite{Shkolnik:2008}. It is suggested that the incidence of these interactions increase with the ratio of the planetary mass to the semi-major axis\cite{Shkolnik:2008}. However, there is no robust evidence for star-planet interactions so far even for 3-5 Jupiter-mass planets on 0.9 to 5-day orbital periods. 55\,Cancri\,e is a 0.02 Jupiter-mass exoplanet in a 0.74-day orbit  and considering the large body of work on this topic\cite{Miller:2015}, we deem unlikely that 55\,Cancri\,e could induce synchronised starspots patterns on its host star. We thus assume in the following that the observed modulation originates from the planet itself.

\begin{figure}
\centering
\includegraphics[width=\textwidth]{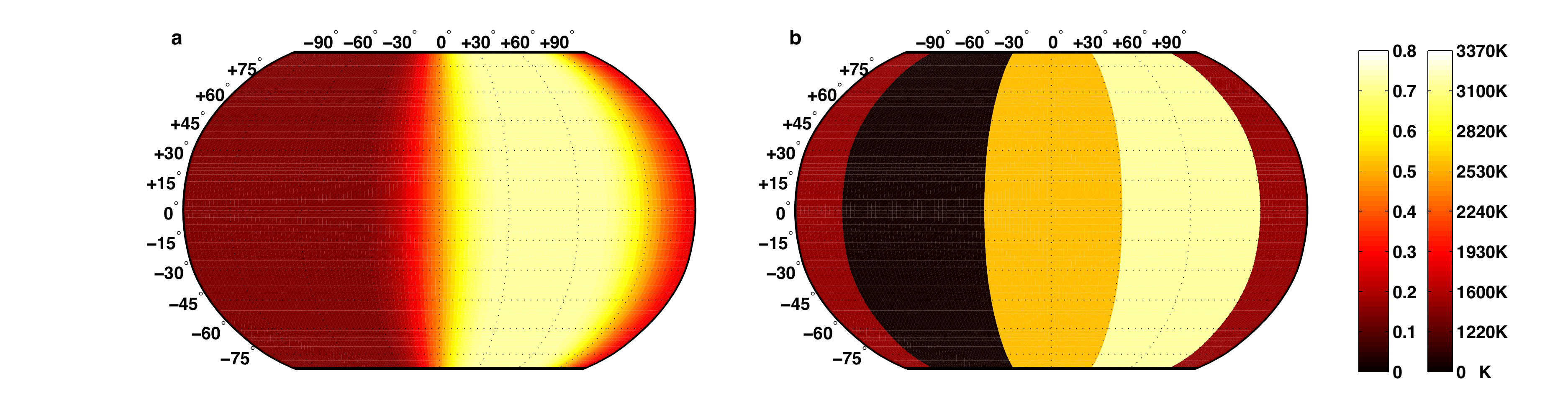}
\caption{\label{fig:map} | {\bf Longitudinal brightness maps of 55\,Cancri\,e.}  Longitudinal brightness distributions as retrieved from the Spitzer/IRAC 4.5$\mu$m phase-curve. The planetary dayside is modelled using two prescriptions. Left: single-band model whose longitude, width and brightness is adjusted in the fit. Right: model including three longitudinal bands whose positions and widths are fixed, but their relative brightness being adjustable. The color-scales indicate the planetary brightness normalised to the stellar average brightness (left) and the corresponding brightness temperature (right) for each longitudinal band.}
\end{figure}

55\,Cancri\,e's phase-curve shape provides constraints on the thermal brightness map of the planet. The phase-curve amplitude translates to a maximum hemisphere averaged brightness temperature of $ 2697^{+268}_{-275} $\,K at 4.5$\mu$m, and a minimum hemisphere averaged brightness temperature of $ 1376^{+344}_{-451} $\,K. We find that the hot spot is centred on the meridian located 41$\pm$12$^{\circ}$ East of the substellar point. We longitudinally map 55\,Cancri\,e's dayside using an MCMC implementation presented in the literature\cite{de-Wit:2012a}. This method has been developed to map exoplanets and to mitigate the degeneracy between the planetary brightness distribution and the system parameters. We model the planetary dayside using two different prescriptions, similar to a previous study\cite{Demory:2013b}. In a first model, we use a single longitudinal band (Figure~\ref{fig:map}, left) with a position and width that are adjusted in the MCMC fit. The second model is similar to the ``beach-ball model''\cite{Cowan:2009b} that uses 3 longitudinal bands with fixed positions and widths (Figure~\ref{fig:map}, right). In both cases, the relative brightness between each longitudinal band is adjusted in the MCMC fit.

The large day-night temperature difference of over 1300\,K indicates the lack of strong atmospheric circulation redistributing energy from the dayside to the nightside of the planet. Such a large contrast could potentially be explained by the extremely high stellar irradiation received on the dayside due to which the radiative timescale may be shorter than the advective timescale, as has been suggested for highly irradiated hot Jupiters which have H$_2$-dominated atmospheres\cite{Showman:2013}. However, the mass, radius, and temperature of the planet are inconsistent with the presence of a significant H$_2$-dominated atmosphere\cite{Gillon:2012,Demory:2015a}, as has also been suggested by the non-detection of H absorption in the Lyman-$\alpha$\cite{Ehrenreich:2012}, though an atmosphere with a higher mean molecular weight cannot be ruled out. It may be possible that a high-mean molecular weight atmosphere in 55\,Cancri\,e, e.g. of H$_2$O or CO$_2$, could also have a lower radiative timescale compared to the advective timescale, thereby explaining the inefficient circulation. However, the observed brightness temperature is unexpectedly high for such an explanation because H$_2$O and CO$_2$ both have significant opacity in the IRAC 4.5 $\mu$m bandpass due to which the upper cooler regions of the atmosphere are probed preferentially. The maximum hemisphere-averaged temperature of 2700\,K $\pm$ 270 K is marginally greater than the highest equilibrium temperature permissible, which is possible for the planetary surface but implausible higher up unless the atmosphere hosts a strong thermal inversion\cite{Madhusudhan:2010}. Alternately, the data may be explained if the planet is devoid of a thick atmosphere of any composition and has a low albedo. Such a hypothesis could explain both the radius of the planet, which is consistent with a purely rocky composition, as well as the lack of strong atmospheric circulation. 

The substantial day-night temperature contrast observed is seemingly incongruous with the observation of a large offset of the hot spot 41$^{\circ}$ East from the substellar point. Such a shift of the hot spot requires efficient energy circulation in the atmosphere\cite{Showman:2013}, contrary to the large day-night contrast observed. An alternate explanation is that the planet harbours an optically thick atmosphere in which heat recirculation takes place but only on the dayside while the gases condense out on the planetary nightside\cite{Heng:2012c}, possibly forming clouds\cite{Demory:2013b}. However, such a scenario requires either the atmosphere to be dominated by vapours of high-temperature refractory compounds, e.g. of silicates\cite{Schaefer:2011,Miguel:2011}, or the nightside temperatures are below freezing for volatiles such as H$_2$O to condense, but the latter is ruled out by our observed nightside temperature of 1380 $\pm$ 400 K. It is possible that there are strong longitudinal inhomogeneities in the chemical composition and emissivity in the atmosphere causing a longitudinally varying optical depth in the 4.5 $\mu$m bandpass that could potentially explain the data. Alternatively, the hot spot offset may be driven by an eastward molten lava flow on the dayside surface of the planet, which would have a viscosity more similar to water at room temperature than to solid rock. At the observed maximum hemisphere-averaged temperature of $\sim$2700\,K silicate-based rocks are expected to be molten\cite{Lutgens:2000}, while the night side temperature of $\sim$1380\,K can be cool enough to sustain a partially to mostly solid surface, where rock viscosities would be several orders of magnitude larger than on the day side.

Additional constraints resulting from the estimated atmospheric escape induced by the nearby host star suggest that it is unlikely that 55\,Cancri\,e is harbouring a thick atmosphere. We find that the surface pressure of 55\,Cancri\,e needs to be larger than 31 kbar in order to survive over the stellar lifetime, which favours an atmosphere-less scenario (see Methods).

From the 3-longitudinal band model fit, we find that the region of maximum thermal emission is located 30 to 60$^{\circ}$ East of the substellar point, with brightness temperatures in excess of 3100\,K. We find that tidal dissipation can explain only a fraction of this reemitted radiation (see Methods), suggesting that an additional, currently unknown source provides a sizeable contribution to the infrared emission of 55\,Cancri\,e.

\newcounter{firstbib}

%%%%%%%%%%%%%%
%%%% ADDENDUM %%%
%%%%%%%%%%%%%%

\begin{addendum}

\item We thank D. Deming, D. Apai and A. Showman for discussions as well as the Spitzer Science Center staff for their assistance in the planning and executing of these observations. This work is based on observations made with the Spitzer Space Telescope, which is operated by the Jet Propulsion Laboratory, California Institute of Technology under a contract with NASA. Support for this work was provided by NASA through an award issued by JPL/Caltech. M.G. is Research Associate at the Belgian Funds for Scientific Research (F.R.S-FNRS). VS was supported by the Simons Foundation (Award Number 338555, VS).

\item[Author Contribution] B.-O.D. initiated and led the {\it Spitzer} observing programme, conducted the data analysis and wrote the paper. M.G performed an independent analysis of the dataset. E.B. carried out the simulations assessing the amplitude of tidal heating in 55\,Cancri\,e's interior. J.d.W. performed the longitudinal mapping of the planet. N.M. wrote the interpretation section with inputs from E.B., K.H., V.S., R.H., N.L. and T.K. J.K., B.B., S.K. and D.Q. contributed to the observing programme. All authors commented on the manuscript.

\item[Author Information] Reprints and permissions information is available at www.nature.com/reprints. The authors declare no competing financial interests. Correspondence and requests for materials should be addressed to B.-O.D (bod21@cam.ac.uk).

\end{addendum}

\clearpage

%%%%%%%%%%%%%%
%%%% FIGURES %%%%
%%%%%%%%%%%%%%

% Table 1 - MCMC parameters
\begin{table*}
\scriptsize
\centering

\begin{tabular}{ll}
\hline
{\bf Planetary basic parameters}\\

Planet/star radius ratio $R_p/R_s$ & $0.0187^{+0.0007}_{-0.0007}$ \\
$b=a \cos i /R_{\star}$ [$R_{\star}$] & $0.41^{+0.05}_{-0.05}$ \\
$T_0 - 2,\!450,\!000$ [BJD$_{\rm TDB}$]& $5733.013^{+0.007}_{-0.007}$ \\
Orbital period $P$ [days] & $0.736539^{+0.000007}_{-0.000007}$ \\
Orbital semi-major axis $a$ [AU]  & $0.01544^{+0.00009}_{-0.00009}$ \\
Orbital inclination $i$ [deg]  & $83.3^{+0.9}_{-0.8}$ \\
Mass $M_{p}$ [$M_{\oplus}$] $^a$ & $8.08^{+0.31}_{-0.31}$ \\
Radius $R_{p}$ [$R_{\oplus}$]  & $1.91^{+0.08}_{-0.08}$ \\
Mean density $\rho_{p}$ [g\, cm$^{-3}$]  &$6.4^{+0.8}_{-0.7}$ \\
Surface gravity $\log g_p$ [cgs]  & $3.33^{+0.04}_{-0.04}$ \\
\hline
{\bf Planetary emission parameters from this work}\\

Phase-curve amplitude [ppm] & $197\pm34$ \\
Phase-curve offset, East [$^{\circ}$] & $41\pm12$\\
Occultation depth (mid-eclipse) [ppm] & $154\pm23$ \\
Maximum hemisphere averaged temperature [K] & $ 2697^{+268}_{-275} $  \\
Minimum hemisphere averaged temperature [K] & $ 1376^{+344}_{-451} $   \\
Average dayside temperature [K] & $ 2349^{+188}_{-193} $   \\
\hline
\end{tabular}

\caption{\label{tab:res} | {\bf 55\,Cancri\,e planetary parameters.} Results from the MCMC combined fit. Values indicated are the median of the posterior distributions and the 1-$\sigma$ corresponding credible intervals. \\
$^a$ Mass prior obtained from the literature\cite{Nelson:2014a}.}
\end{table*}

\clearpage

%%%%%%%%%%%%%%
%%%% METHODS %%%%
%%%%%%%%%%%%%%

\setcounter{figure}{0}
\setcounter{table}{0}
\renewcommand{\tablename}{Extended Data Table}
\renewcommand{\figurename}{Extended Data Figure}

\begin{figure}
\centering
\includegraphics[width=\textwidth]{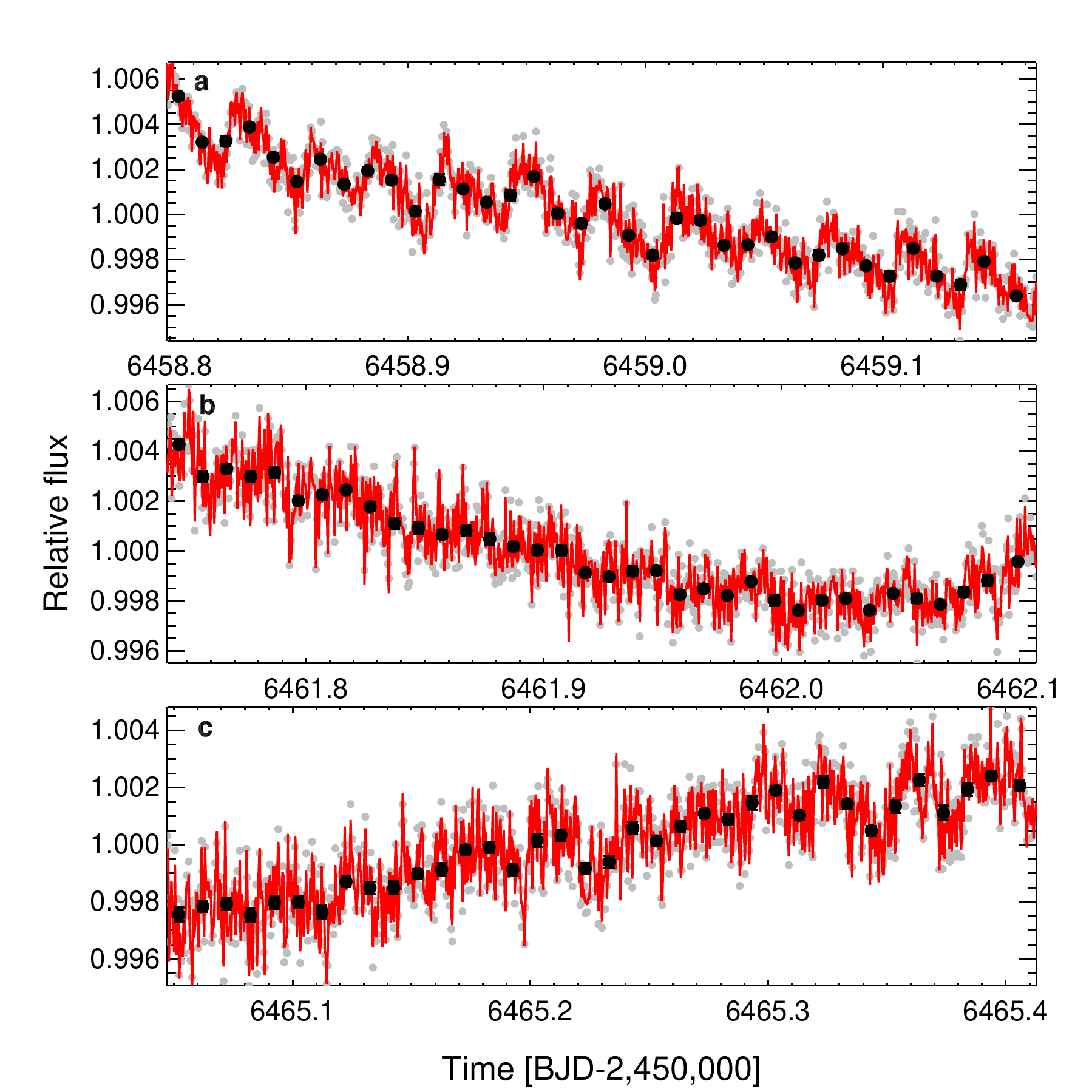}
\caption{\label{fig:raw1} | {\bf 55\,Cancri\,e raw photometry.} The raw data for time-series acquired on 2013 15/06 (a), 18/06 (b) and 21/06 (c) are shown. The best-fit instrumental+astrophysical model is superimposed in red. Grey filled circles are data binned per 30s. Black filled circles are data binned per 15 minutes. The error bars are the standard deviation of the mean within each time bin.}
\end{figure}

\begin{figure}
\centering
\includegraphics[width=\textwidth]{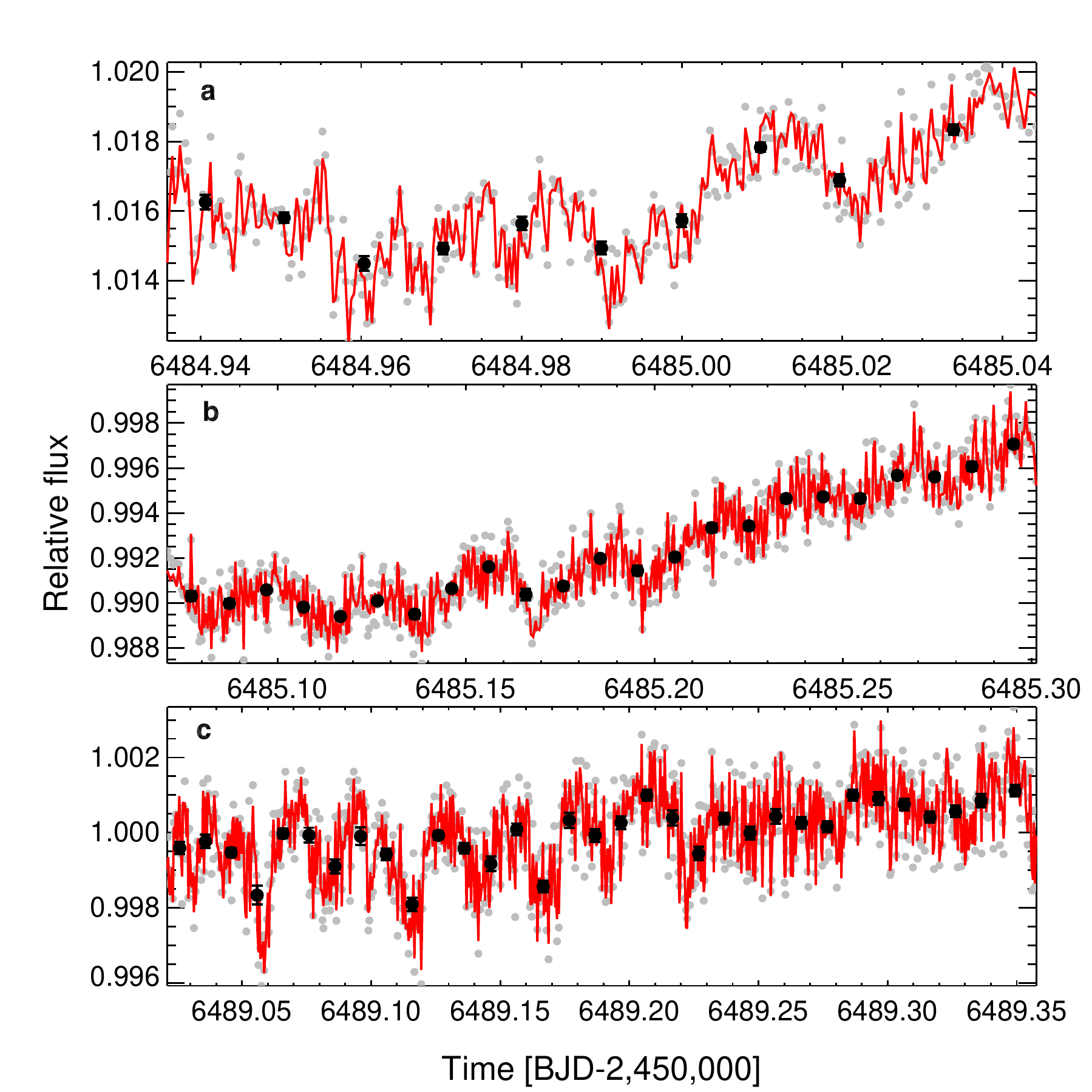}
\caption{\label{fig:raw3} | {\bf 55\,Cancri\,e raw photometry.} The raw data for time-series acquired on 2013 11/07 (a and b) and 15/07 (c) are shown. The best-fit instrumental+astrophysical model is superimposed in red. Grey filled circles are data binned per 30s. Black filled circles are data binned per 15 minutes. The error bars are the standard deviation of the mean within each time bin.}
\end{figure}

\begin{figure}
\centering
\includegraphics[width=\textwidth]{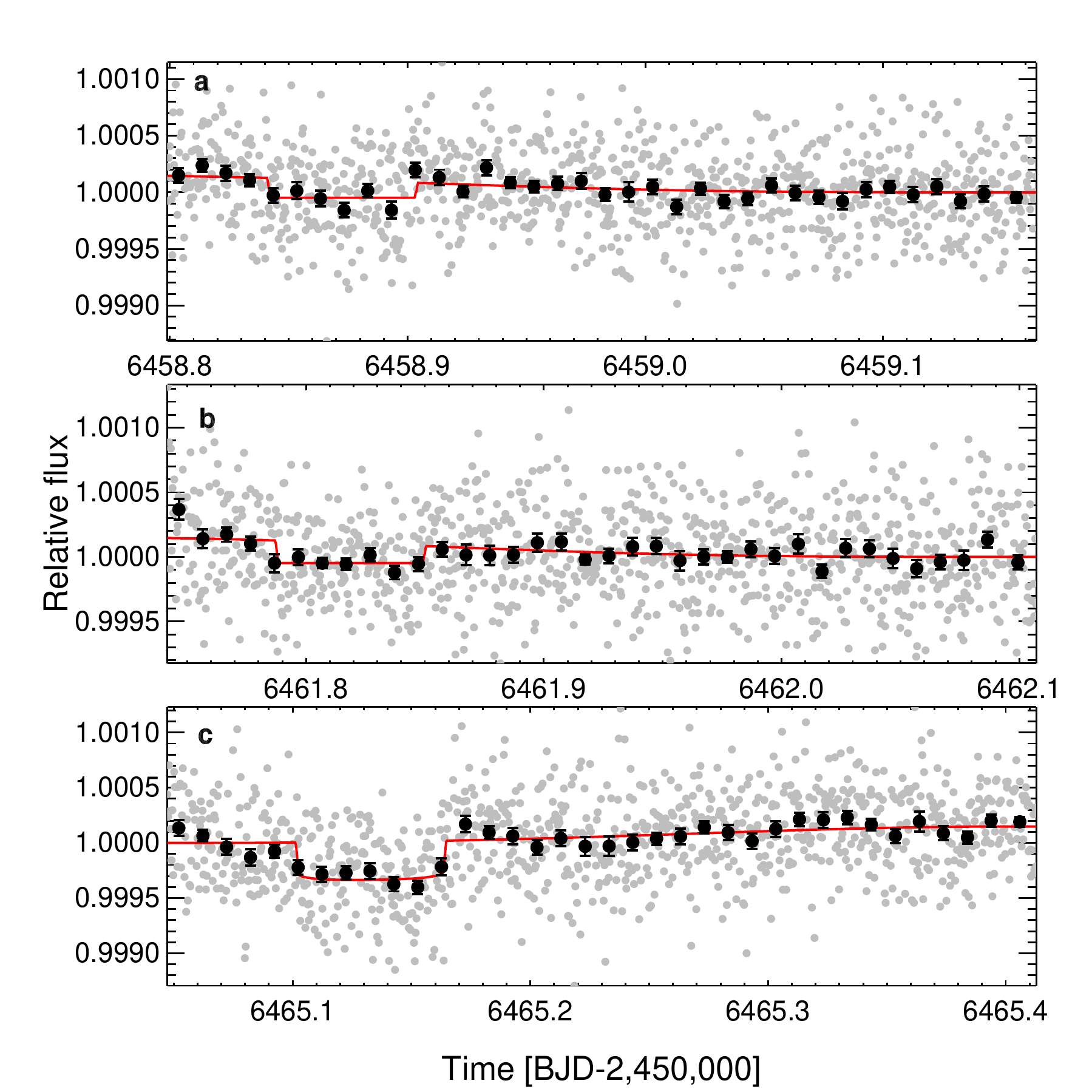}
\caption{\label{fig:cor1} | {\bf 55\,Cancri\,e corrected photometry.} The detrended data for time-series acquired on 2013 15/06 (a), 18/06 (b) and 21/06 (c) are shown. The best-fit instrumental+astrophysical model is superimposed in red. Grey filled circles are data binned per 30s. Black filled circles are data binned per 15 minutes. The error bars are the standard deviation of the mean within each time bin.}
\end{figure}

\begin{figure}
\centering
\includegraphics[width=\textwidth]{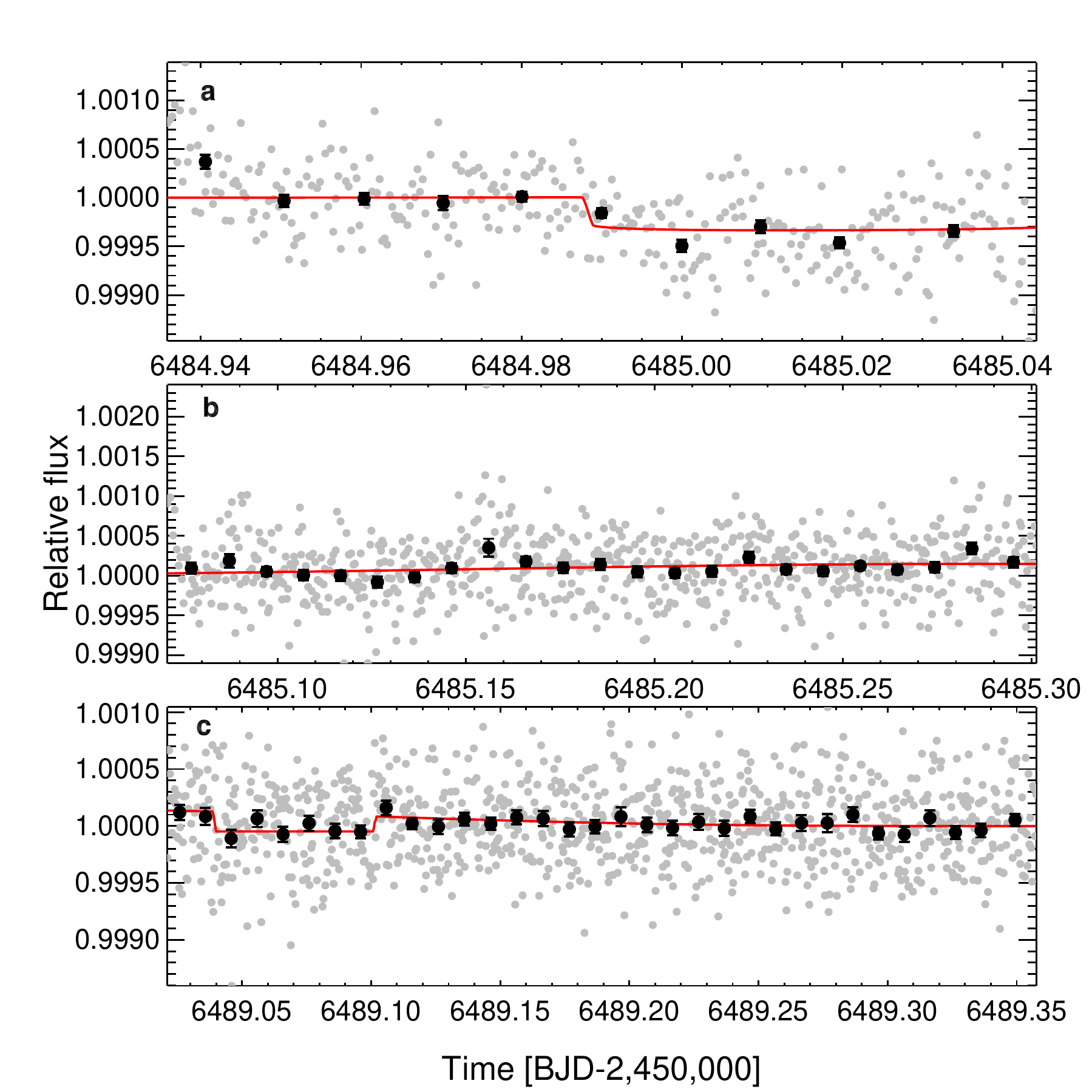}
\caption{\label{fig:cor3} | {\bf 55\,Cancri\,e corrected photometry.} The detrended data for time-series acquired on 2013 11/07 (a and b) and 15/07 (c) are shown. The best-fit instrumental+astrophysical model is superimposed in red. Grey filled circles are data binned per 30s. Black filled circles are data binned per 15 minutes. The error bars are the standard deviation of the mean within each time bin.}
\end{figure}

\begin{methods}

\subsection{Data reduction.} 

We observed four phase-curves of 55\,Cancri\,e with the Spitzer Space Telescope in the IRAC/4.5-$\mu$m channel as part of our program ID 90208. Because of downlink constraints, these four phase-curve observations were split in 8 separate observations (or Astronomical Observation Request, AOR) lasting each half an orbit of 55\,Cancri\,e. Details for each AOR can be found in Extended Data Table~\ref{tab:obs}. The corresponding data can be accessed using the Spitzer Heritage Archive (\url{http://sha.ipac.caltech.edu}). All AORs have been acquired in stare mode using a consistent exposure time (0.02 s). All of our data have been obtained using the PCRS peak-up mode, which allows the observer to place the target on a precise location on the detector to mitigate the intra-pixel sensitivity variations. This observing mode increases the pointing stability and reduces the level of correlated noise in the data by a factor of 2-3\cite{Ballard:2014}. The AOR 48072960 experienced a 30-min interruption during data acquisition, which forced us to treat both parts of that AOR separately in the rest of this section. We do not retain the 30-min long PCRS sequences in our analysis since the motion of the star on the detector yields large correlated noise in these datasets.
Our reduction uses the basic calibrated data (BCD) that are downloaded from the Spitzer Archive. BCD are FITS data cubes comprising 64 frames of 32x32 pixels each. Our data reduction code reads each frame, converts fluxes from the {\it Spitzer} units of specific intensity (MJy/sr) to photon counts, and transform the data timestamps from BJD$_{\rm UTC}$ to BJD$_{\rm TDB}$ using existing procedures\cite{Eastman:2010a}. We do not find necessary to discard specific subarray frames. During the reduction process, we compute the flux, position and FWHM in each of the 64 frames of each datacube and check for any discrepant value in these three parameters. The frames that have any of these parameters discrepant by more than 5-sigma are discarded. The centroid position on the detector is determined by fitting a Gaussian to the marginal X, Y distributions using the {\sl GCNTRD} procedure part of the IDL Astronomy User's Library\cite{Landsman:1993}. We also fit a two-dimensional Gaussian to the stellar PRF following previous studies\cite{Agol:2010}. We find that determining the centroid position using {\sl GCNTRD} results in a smaller dispersion of the fitted residuals by 10 to 15\% across our dataset, in agreement with other {\it Spitzer} analyses\cite{Beerer:2011}. We then perform aperture photometry for each dataset using a modified version of the {\sl APER} procedure using aperture sizes ranging from 2.2 to 4.4 pixels in 0.2 pixels intervals. We choose the optimal aperture size based on the minimisation of ${\rm RMS}\times \beta_{\rm red}^2$ for each AOR, where $\beta_{\rm red}$ is the red-noise contribution\cite{Gillon:2012a}. The red noise is assessed over 60-min timescales as smaller timespans are irrelevant for the phase-curve signal whose periodicity is 18 hours. We measure the background contribution on each frame using an annulus located 10 to 14 pixels from the centroid position. Our code also determines the PRF's FWHM along the X and Y axes. We use a moving average based on forty consecutive frames to discard datapoints that are discrepant by more than 5$\sigma$ in background, $x$-$y$ position or FWHM. We find that in average 0.06\% of the datapoints are discarded. The resulting time-series are finally combined per 30-s bins to speed up the analysis, which has shown to have no incidence on the system parameters values and uncertainties\cite{Demory:2015a}. We show the retained aperture size, corresponding RMS and $\beta_{\rm red}$ for each dataset in Extended Data Table~\ref{tab:obs}.

\subsection{Photometric Analysis.} 

\subsubsection*{\--- Intra-pixel sensitivity correction.}
\label{ips}
We use an implementation of the BLISS (BiLinearly-Interpolated Sub-pixel Sensitivity)\cite{Stevenson:2012a} to account for the intra-pixel sensitivity variations, similarly to a previous study using the same dataset\cite{Demory:2015a}.

The BLISS algorithm uses a bilinear interpolation of the measured fluxes to build a pixel-sensitivity map. The data are thus self-calibrated. Our implementation of this algorithm is included in the Markov Chain Monte Carlo (MCMC) framework already presented in the literature\cite{Gillon:2012a}. The improvement brought by any pixel-mapping technique such as BLISS requires that the stellar centroid remains in a relatively confined area on the detector, which warrants an efficient sampling of the $x$/$y$ region, thus an accurate pixel map. In our implementation of the method, we build a sub-pixel mesh made of $n^2$ grid points, evenly distributed along the $x$ and $y$ axes. The BLISS algorithm is applied at each step of the MCMC fit. The number of grid points is determined at the beginning of the MCMC by ensuring that at least 5 valid photometric measurements are located in each mesh box. Similar to two recent studies\cite{Lanotte:2014, Demory:2015a}, we find that a further reduction of the level of correlated noise in the photometry is brought by the inclusion of the PRF's FWHM along the $x$ and $y$ axes in the baseline model. The PRF evolves with time and its properties are not accounted for by the BLISS algorithm. We thus combine the BLISS algorithm to a linear function of the PRF's FWHM along both the $x$ and $y$ axes. In addition, the baseline model for each AOR includes a flux constant. We find that including a model of the PRF's FWHM decreases the Bayesian Information Criterion (BIC\cite{Schwarz:1978}) by $\Delta_{BIC}=591$. We show the raw datasets with the best fit instrumental+astrophysical model superimposed in red on Extended Data Figures~\ref{fig:raw1} \--- \ref{fig:raw3}. The corrected photometry is shown on Extended Data Figures~\ref{fig:cor1} \--- \ref{fig:cor3}. The phase-curve modulation is clearly noticeable in each AOR. The behaviour of the RMS vs. binning is shown for each dataset on Extended Data Figure~\ref{fig:rms}.

\subsubsection*{\--- Model Comparison}

In our first MCMC analysis, we use the following functional form to model the planet infrared emission variation:

\begin{equation}
F = F_P + Tr + Oc
\end{equation}

where $F$ is the observed flux, $F_P$ is the phase modulation driven by the planet, $Tr$ is the transit model and $Oc$ is the occulation model.

We use a Lambertian\cite{Sobolev:1975} functional form for $F_P$:

\begin{equation}
F_P = A_{\rm phase} \frac{\sin z + (\pi-z) \cos z}{\pi}
\label{eq:ph0}
\end{equation}

$A_{\rm phase}$ being the phase amplitude and with

\begin{equation}
\cos z = -\sin i \cos[2\pi(\phi + \theta_{phase})]
\label{eq:ph1}
\end{equation}

\begin{equation}
\phi = \frac{2 \pi}{P} (t-T_0) 
\label{eq:ph1b}
\end{equation}

where $\theta_{phase}$ is the phase-curve offset, $P$ the orbital period and $T_0$ the transit centre.

The transit lightcurve model $MA$, published elsewhere\cite{Mandel:2002} is summarised as:

\begin{equation}
Tr = MA(dF_{\rm tr}, P, b, M_{\star}, c_1, c_2, t)
\label{eq:ph2}
\end{equation}

\begin{equation}
Oc = MA(dF_{\rm occ}, P, b, M_{\star}, t)
\label{eq:ph2b}
\end{equation}

for the transit ($Tr$) and occultation ($Oc$), where $dF_{\rm tr}$ and $dF_{\rm occ}$ are respectively the transit and occultation depths, $b$ the impact parameter, the limb-darkening linear combinations $c_1=2u_1+u_2$ and $c_2=u_1-2u_2$, where $u_1$ and $u_2$ are the quadratic coefficients drawn from theoretical tables\cite{Claret:2011} using published stellar parameters\cite{von-Braun:2011a}.

We have also experimented using a sinusoid functional form for the phase variation:

\begin{equation}
F_P = A_{\rm phase} \cos (\phi + \theta_{phase})
\label{eq:ph3}
\end{equation}

The fit using a sinusoid results in an amplitude of $218\pm50$ ppm and an offset value of $68\pm24^{\circ}$ East of the substellar point, in agreement with our results using a Lambertian functional form ($197\pm34$ ppm and $41\pm12^{\circ}$).

We notice that a Lambertian sphere model provides a better fit to the data than a sine function, with a $\Delta_{\rm BIC}=11$.

We also perform another MCMC analysis with no phase-curve model, hence removing two degrees of freedom (phase amplitude and phase offset). We find a $\Delta_{\rm BIC}=21$ in favour of the model including the phase-curve model.

We also run an MCMC fit that includes the phase amplitude alone but not the phase offset. We find that this fit produces only a marginal $\chi^2$ improvement over the MCMC fit with no phase-curve model but this improvement is penalised by the extra degree of freedom according to the BIC. We indeed obtain a $\Delta_{\rm BIC}=25$ in favour of the model including the phase-curve offset.

Altogether, this model comparison thus confirms that a phase-curve model that includes a phase offset is the favoured functional form according to the BIC.
 
\subsubsection*{\--- Additional analyses.}

We conduct two additional analyses of our entire dataset in order to assess the robustness of our initial detection that used the BLISS mapping technique. In these two analyses, we use different approaches to 1) model the detector's intra-pixel sensitivity and 2) change the input data format. In a first analysis, we use a simple polynomial detrending approach with a functional form including the centroid position (fourth order) and FWHM (first order) alone. We have experimented different polynomial orders (from 1 to 4) for these two parameters and found this combination to globally minimise the BIC. Each AOR has its own set of baseline coefficients. As for the BLISS mapping, the polynomial detrending is included in the MCMC fit so the baseline model and the system parameters are adjusted simultaneously to efficiently propagate the uncertainties to the final parameters. We find a level of correlated noise in the data that is only slightly larger (about 10\%) than the one obtained with the BLISS mapping technique. Using this method we find a phase-curve minimum of 36$\pm$41 ppm, a maximum of 187$\pm$41 ppm and offset of 50$\pm$13 degrees East, compared to the 47$\pm$34 ppm, 197$\pm$34 ppm and 41$\pm$12 degrees East values from our analysis using the BLISS mapping. As previously shown\cite{Lanotte:2014}, the addition of the PRF's FWHM in the baseline model significantly improves a fit based on a centroid position alone and most importantly, it allows an acceptable fit on time-series that are 8-hour long. 

In a second analysis, we aim at assessing whether the phase-curve signal persists when we split our input data. All of our AORs have a duration of nearly 9 hours and we elect to split each of them in two to reduce the duration of each individual dataset to 4.5 hours. The functional forms of the baseline models are the same as for the analysis employing the unsegmented input data described above. In this additional test, we find a phase-curve minimum of 51$\pm$51ppm, a maximum of 216$\pm$51ppm, and a phase-curve offset of 54$\pm$16 degrees East of the substellar point. These results are in good agreement with our main analysis. The uncertainties on the phase-curve parameters are larger in this case because of the time-series segmentation, which does not constrain the baseline coefficients as effectively as for longer datasets. The phase-curves obtained from these additional analyses are shown on Extended Data Figure~\ref{fig:poly}.

We finally note that the phase-curve peak is located close to the start of half of our observations and toward the end of the other half datasets, which was necessary due to {\it Spitzer} downlink limitations. We deem this pattern being purely coincidental for two reasons. Firstly, if our reported phase curve was due to uncorrected systematics, it would be unlikely that the systematics would produce an upward trend in half of the data and a downward slope in the other half. These datasets are independent between them and there is no relationship between those obtained from transit to occultation and the other obtained from occultation to transit. There is also no correlation with the centroid position on the detector. Secondly, if the phase peak offset was happening after or before this discontinuity, it would have been clearly detected in the continuous parts of our dataset. But instead only gradual slopes are seen in both datasets. A comparison with data obtained in the same year with the Microvariability and Oscillations of STars (MOST) satellite (D. Dragomir, personal communication) shows an agreement in the phase-curve amplitude and offset values derived from both facilities.

\subsection{Longitudinal Mapping.}

The key features of 55\,Cancri\,e's phase-curve translate directly into constraints on maps\cite{Knutson:2007, Cowan:2008} assuming a tidally-locked planet on a circular orbit. A planetary phase-curve $\frac{F_p}{F_{\star}}$ measures the planetary hemisphere-averaged relative brightness $\frac{<I_p>}{<I_{\star}>}$ as follows:
\begin{equation}
\frac{F_p}{F_{\star}}(\alpha)=\frac{<I_p>(\alpha)}{<I_{\star}>}\left(\frac{R_p}{R_{\star}}\right)^2  
\label{eq:lm}
\end{equation}
where $\alpha$ is the orbital phase.

The longitudinal mapping technique used here\cite{de-Wit:2012a} aims at mitigating the degeneracy between the planetary thermal brightness distribution and the system parameters. This part of the analysis is independent from the light curve analysis presented above. Therefore, we fix here the system parameters to the ones derived from a previous study\cite{Demory:2015a}, which is based on the entire 55\,Cancri\,e's {\it Spitzer} dataset. We note that using this prior information for the purpose of longitudinal mapping is adequate since the degeneracy between the planetary brightness distribution and system parameters is only significant in the context of eclipse mapping\cite{de-Wit:2012a}. We follow here the same approach as for Kepler-7b\cite{Demory:2013b} and use two families of models, similar to the ``beachball models'' introduced by a previous study\cite{Cowan:2009b}: one using $n$ longitudinal bands with fixed positions on the dayside and another using longitudinal bands whose positions and widths are jump parameters in the MCMC fit. We choose a 3-fixed-band model and 1-free-band model so as to extract both 55\,Cancri\,e's longitudinal dependence of the dayside brightness and the extent of the ``bright'' area. Increasing $n$ to 5 yields a larger BIC than for $n=3$.  For both models, we compute each band's amplitude from their simulated lightcurve by using a perturbed singular value decomposition method. The 1-free-band model (Fig.~\ref{fig:map}, left) finds a uniformly bright longitudinal area extending from $5\pm18^{\circ}$ west to $85\pm18^{\circ}$ east with a relative brightness $0.72\pm0.18$, compared to a brightness of $0.15\pm0.05$ for the rest of the planet. The 3-fixed-band model finds bands of relative brightness decreasing from the west to the east with the following values: $<0.21$ (3-$\sigma$ upper limit), $0.58\pm0.15$, $0.74\pm0.15$ compared to the night-side contribution of $0.17\pm0.06$.

\subsection{55\,Cancri\,e's thermal emission variability.}

Earlier this year, a study\cite{Demory:2015a} has been published reporting variability in 55\,Cancri\,e's thermal emission between 2012 and 2013, from occultation measurements. Several tests regarding the robustness of the variability pattern were conducted, including three different analyses, using the BLISS mapping, polynomial detrending and the recently published pixel-level decorrelation method\cite{Deming:2015}. These three approaches confirmed the variability of the thermal emission of the planet between 2012 and 2013 with similar uncertainties. We thus consider very likely that the planet emission is varying, but on timescales that are significantly longer than the timespan of the 2013 observations alone (just a month) used in the present paper. No variability is reported in the 2013 data alone\cite{Demory:2015a}. These factors comforted us in combining the 2013 observations together and in using a single phase-curve model. Furthermore, we detect the phase curve shape in all individual datasets in addition to the combined phase-folded time-series. This strengthens our conclusion that it is unlikely that stellar variability would build the combined phase curve shape from individual stellar events taken at different times over the month of observations.

\subsection{Brightness temperatures.}

We use an observed infrared spectrum of 55\,Cancri\,e\cite{Crossfield:2012d} to compute the brightness temperatures in the IRAC's 4.5$\mu$m bandpass from the $\frac{F_P}{F_{\star}}$ values derived from the MCMC fits.

\subsection{Constraints on 55\,Cancri\,e's atmosphere.} 

If an atmosphere was present, the large temperature contrast between the dayside and nightside hemispheres suggests that the radiative cooling time ($t_{\rm rad}$) is less than the dynamical time scale ($t_{\rm dyn}$), resulting in a poor redistribution of heat from the dayside to the nightside.  This sets a constraint on the mean molecular weight, which we may estimate.  The zonal velocity is $v \sim \sqrt{{\cal R} \Delta T} \sim 1$ km s$^{-1}$, where ${\cal R}$ is the specific gas constant, $\Delta T = 1460$ K is the temperature difference between the hemispheres and we have ignored an order-of-unity correction factor associated with the pressure difference between the hemispheres\cite{Menou:2012b}.  If we enforce $t_{\rm rad} < t_{\rm dyn}$, then we obtain
\begin{equation}
\mu > \frac{ {\cal R}_{\rm univ} \left( \Delta T \right)^{1/3}}{T^2_{\rm day}} \left( \frac{P_{\rm day}}{\sigma_{\rm SB} R \kappa g}\right)^{2/3},
\end{equation}
where ${\cal R}_{\rm univ} = 8.3144598 \times 10^7$ erg K$^{-1}$ g$^{-1}$ is the universal gas constant, $T_{\rm day} = 2700 K$ is the dayside temperature, $\sigma_{\rm SB}$ is the Stefan-Boltzmann constant, $R=1.91 R_\oplus$ is the planetary radius, $\kappa = 2/7$ is the adiabatic coefficient and $g = 10^{3.33}$ cm s$^{-2}$ is the surface gravity.  The dayside pressure, $P_{\rm day}$, is the only unknown parameter in the preceding expression.  If we set $P_{\rm day} = 1$ bar, then we obtain $\mu > 9$.  This estimate further suggests that a hydrogen-dominated atmosphere is unlikely and sets a lower limit on the mean molecular weight.

It is unlikely that 55\,Cancri\,e is harbouring a thick atmosphere due to its proximity to its star.  If we assume energy-limited escape\cite{Owen:2013}, then the atmosphere needs to have sufficient mass in order to survive for the stellar age, which translates into a lower limit on the required surface pressure,
\begin{equation}
P > \frac{{\cal L}_{\rm X} R t_{\star} g}{16 \pi G M a^2},
\end{equation}
where ${\cal L}_{\rm X}$ is the X-ray luminosity of the star, $t_\star$ is the stellar age, $G$ is Newton's gravitational constant, $M=8.08 M_\oplus$ is the planetary mass and $a=0.01544$ AU is the semi-major axis.  If we use ${\cal L}_{\rm X} = 4 \times 10^{26}$ erg s$^{-1}$ \cite{Ehrenreich:2012} and $t_\star = 8$ Gyr, then we obtain $P > 31$ kbar.  In other words, the surface pressure of 55\,Cancri\,e needs to be larger than 31 kbar in order to survive atmospheric escape over the stellar lifetime.  Despite the uncertainties associated with estimating the mass loss due to atmospheric escape, this estimate is conservative because the star probably emitted higher X-ray luminosities in the past.  Our suggestion of an atmosphere-less 55\,Cancri\,e is consistent with the trends predicted for super Earths\cite{Owen:2013}.

Hence, it is unlikely that the large infrared peak offset is due to an atmosphere rich with volatiles.  It is more likely that the infrared phase curve of 55\,Cancri\,e is probing non-uniformities associated with its molten rocky surface.

\subsection{Tidal Heating.}

As 55\,Cancri is a multiple planet system, the eccentricity and obliquity of planet e are excited due to the presence of the outer planets. This creates a tidal heat flux which is responsible in part for the planet's thermal emission. In order to evaluate the tidal heat flux contribution, we investigate with N-body simulations (using Mercury-T\cite{Bolmont:2015}) the possible values of the eccentricity and obliquity of planet e for different tidal dissipation. We use the orbital elements and masses for the four outer planets published in a previous study\cite{Nelson:2014a} and the most recent values\cite{Demory:2015a} for the mass, radius and semi-major axis of 55\,Cancri\,e. 

We find that the obliquity of planet $e$ is very low ($<1^{\circ}$) and that the eccentricity is around $10^{-3}$ for the eight orders of magnitude we consider for planet $e$'s tidal dissipation (from $10^{-5}$ to $10$ times Earth's dissipation $\sigma_{\oplus}$). The corresponding tidal heat flux $\phi_{\rm tides}$, or tidal temperature $\phi_{\rm tides}/\sigma)^{1/4}$ increase with the dissipation in the planet: from $10^{-3}$ W/m$^2$ (a few K) to $10^{6}$ W/m$^2$ ($\sim$2000 K). We calculate the occultation depth at 4.5$\mu$m for a range of eccentricities and albedos (0.0 to 1.0) in order to enable a comparison with the dynamical simulations output (Extended Data Figure~\ref{fig:tides}). We find that a combination between large dissipation (10$\sigma_{\oplus}$), eccentricity and obliquity can explain the level of thermal emission observed in 2013, however these solutions do not allow us to reproduce the nightside temperature. In our configuration (no heat re-distribution and assuming an isotropic tidal heat flux), tides do not match our measurements, so an additional heat source is likely responsible for at least part of the large planetary thermal emission observed in 2013.
\end{methods}

%%%%%%%%%%%%%%
%%% METHODS BIB %%%
%%%%%%%%%%%%%%

%%%%%%%%%%%%%%
%%%% SUPP FIG %%%%
%%%%%%%%%%%%%%

\begin{table*}
\scriptsize
\centering
\begin{tabular}{lllllllll}

\hline
Date [UT] & Program ID & AOR \# & AOR duration [h] &  Phase range & Aperture [pix] & Interp. $n$ & RMS/30s [ppm] & $\beta_r$ \\
\hline

2013-06-15 & 90208 & 48070144 & 8.8 		& 0.40 - 0.89 &  2.6 	& 64 & 341 & 1.17 \\
2013-06-18 & 90208 & 48073216 & 8.8 		& 0.39 - 0.89 & 3.0 	& 64 & 340 & 1.00 \\
2013-06-21 & 90208 & 48070656 & 8.8 		& 0.88 - 0.38 & 2.8 	& 70 & 363 & 1.00 \\
2013-06-29 & 90208 & 48073472 & 8.8 		& 0.39 - 0.89 & 3.0 	& 58 & 354 & 1.16 \\
2013-07-03 & 90208 & 48072448 & 8.8 		& 0.90 - 0.39 & 3.2 	& 69 & 376 & 1.34 \\
2013-07-08 & 90208 & 48072704 & 8.0 		& 0.94 - 0.39 & 2.6 	& 77 & 370 & 1.03 \\
2013-07-11 & 90208 & 48072960, p1 & 2.6	& 0.88 - 0.03 &  2.6 	& 22 & 388 & 1.77 \\
2013-07-11 & 90208 & 48072960, p2 & 5.5    	& 0.07 - 0.38 &  2.6 	& 58 & 389 & 1.83 \\
2013-07-15 & 90208 & 48073728 & 8.1 		& 0.43 - 0.89 & 3.4 	& 73 & 348 & 1.00 \\
\hline
\end{tabular} 

\caption{\label{tab:obs} | {\bf 55\,Cancri\,e Spitzer dataset.} Astronomical Observation Request (AOR) properties for the Spitzer/IRAC 4.5-$\mu$m data used in the present study. This table also indicates the planetary orbital phase covered by each AOR as well as the number of interpolation points relevant to the BLISS algorithm.}

\end{table*}

\begin{figure}
\centering
\includegraphics[width=\textwidth]{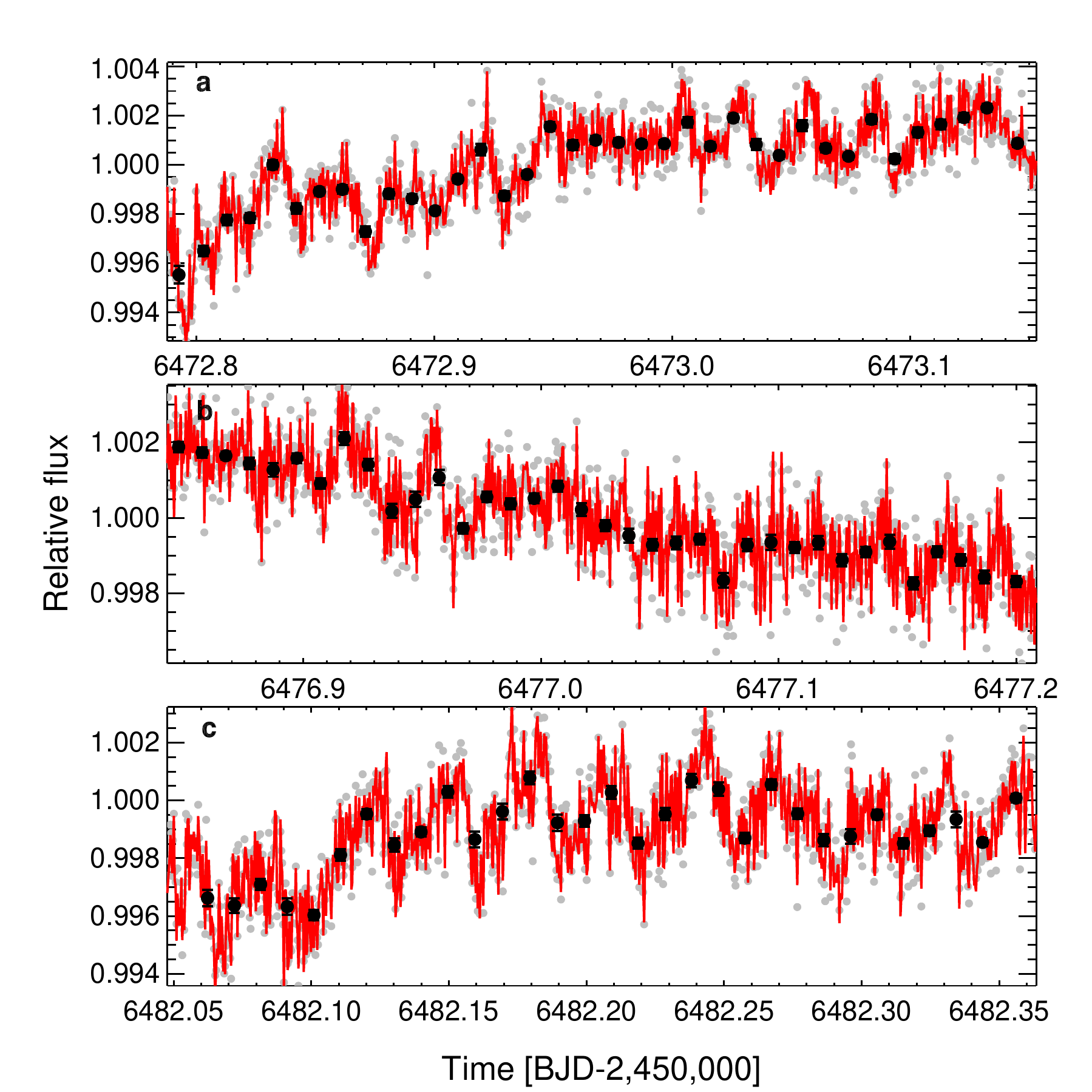}
\caption{\label{fig:raw2} | {\bf 55\,Cancri\,e raw photometry.} The raw data for time-series acquired on 2013 29/06 (a), 03/07 (b) and 08/07 (c) are shown. The best-fit instrumental+astrophysical model is superimposed in red. Grey filled circles are data binned per 30s. Black filled circles are data binned per 15 minutes. The error bars are the standard deviation of the mean within each time bin.}
\end{figure}

\begin{figure}
\centering
\includegraphics[width=\textwidth]{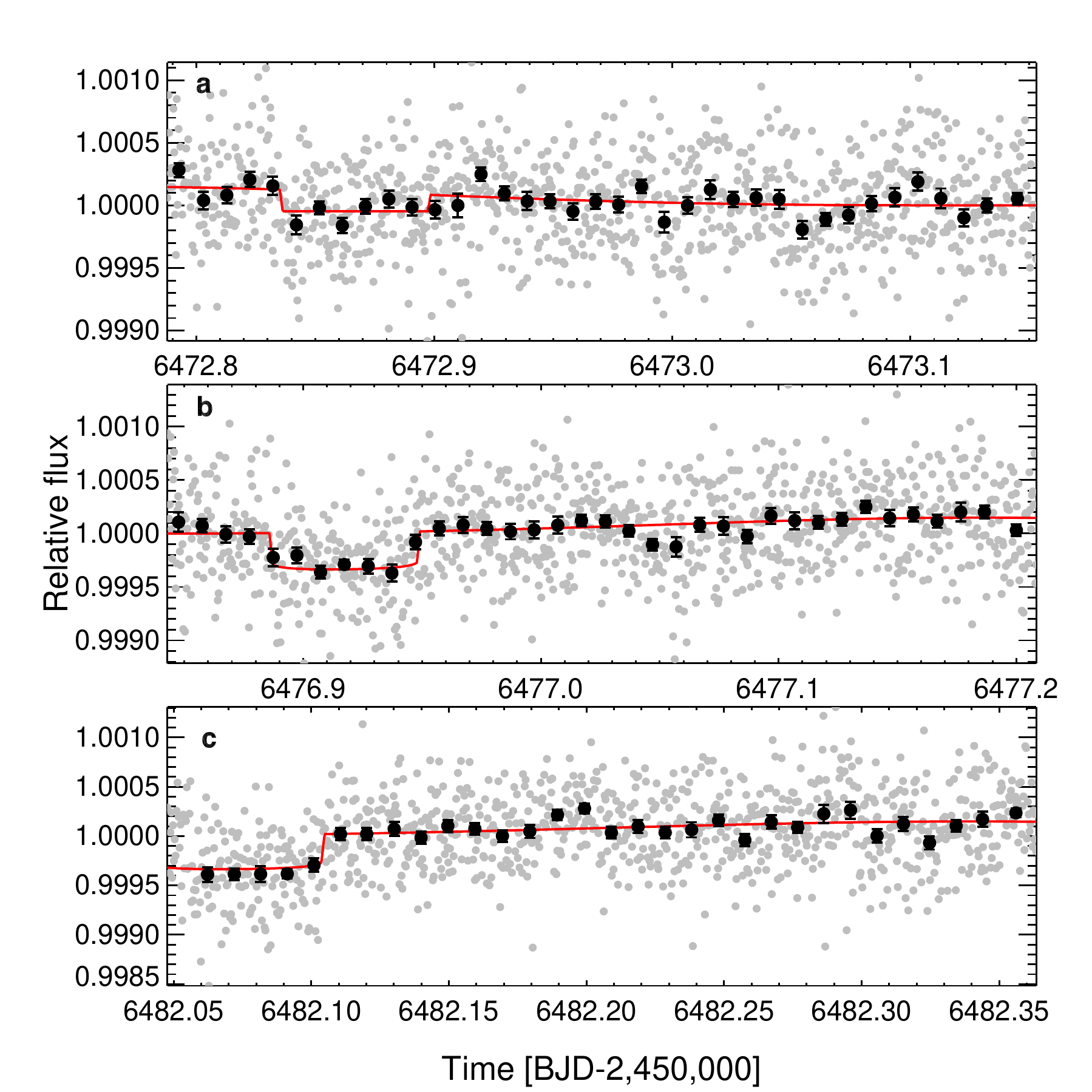}
\caption{\label{fig:cor2} | {\bf 55\,Cancri\,e corrected photometry.} The detrended data for time-series acquired on 2013 29/06 (a), 03/07 (b) and 08/07 (c) are shown. The best-fit instrumental+astrophysical model is superimposed in red. Grey filled circles are data binned per 30s. Black filled circles are data binned per 15 minutes. The error bars are the standard deviation of the mean within each time bin.}
\end{figure}

\begin{figure}
\centering
\includegraphics[width=0.8\textwidth]{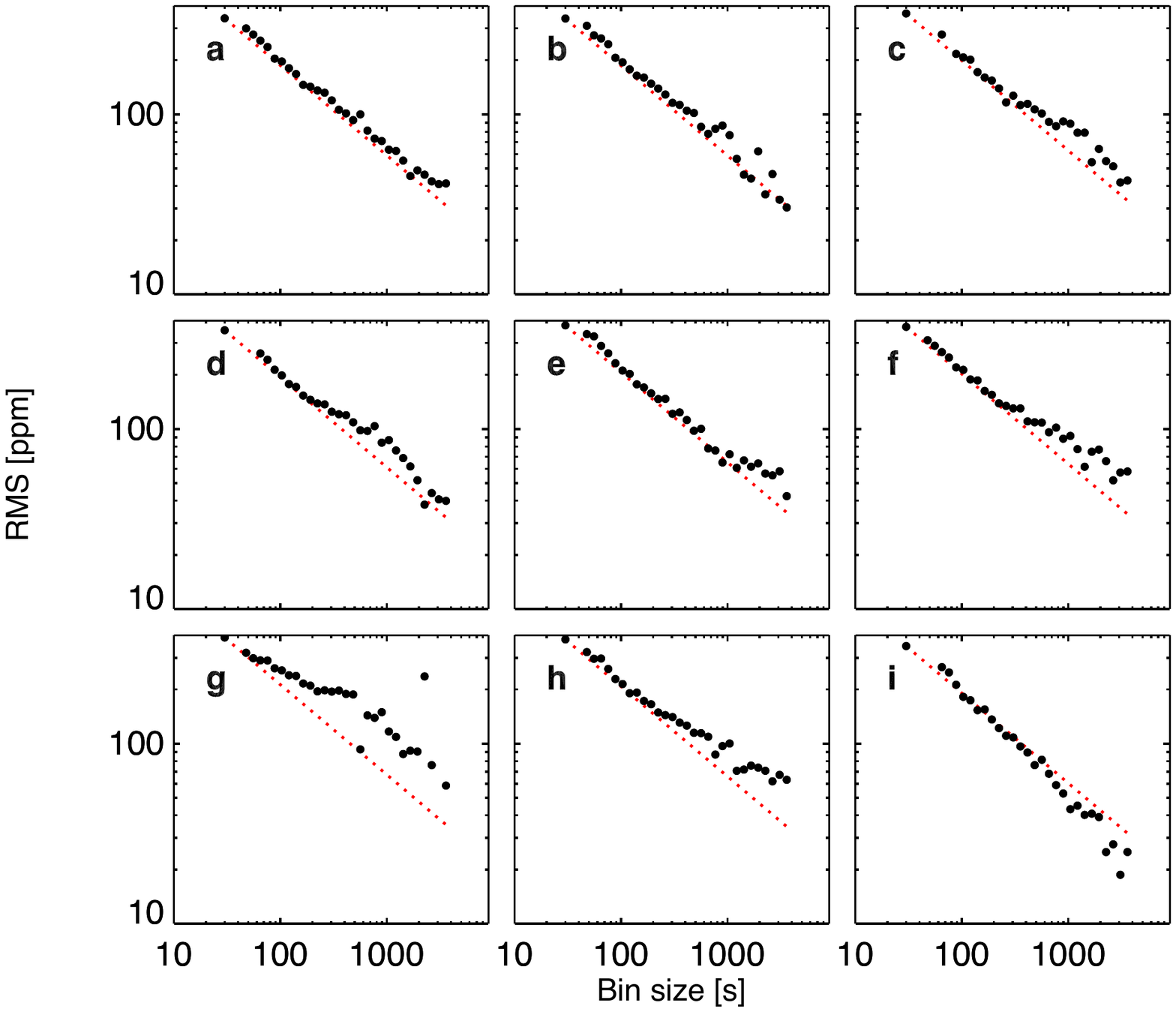}
\caption{\label{fig:rms} | {\bf Photometric RMS vs. binning for all datasets.} Black filled circles indicate the photometric residual RMS for different time bins. Each panel corresponds to each individual dataset (a to i, increasing observing date).  The expected decrease in Poisson noise normalised to an individual bin (30s) precision is shown as a red dotted line. }
\end{figure}

\begin{figure}
\centering
\includegraphics[width=\textwidth]{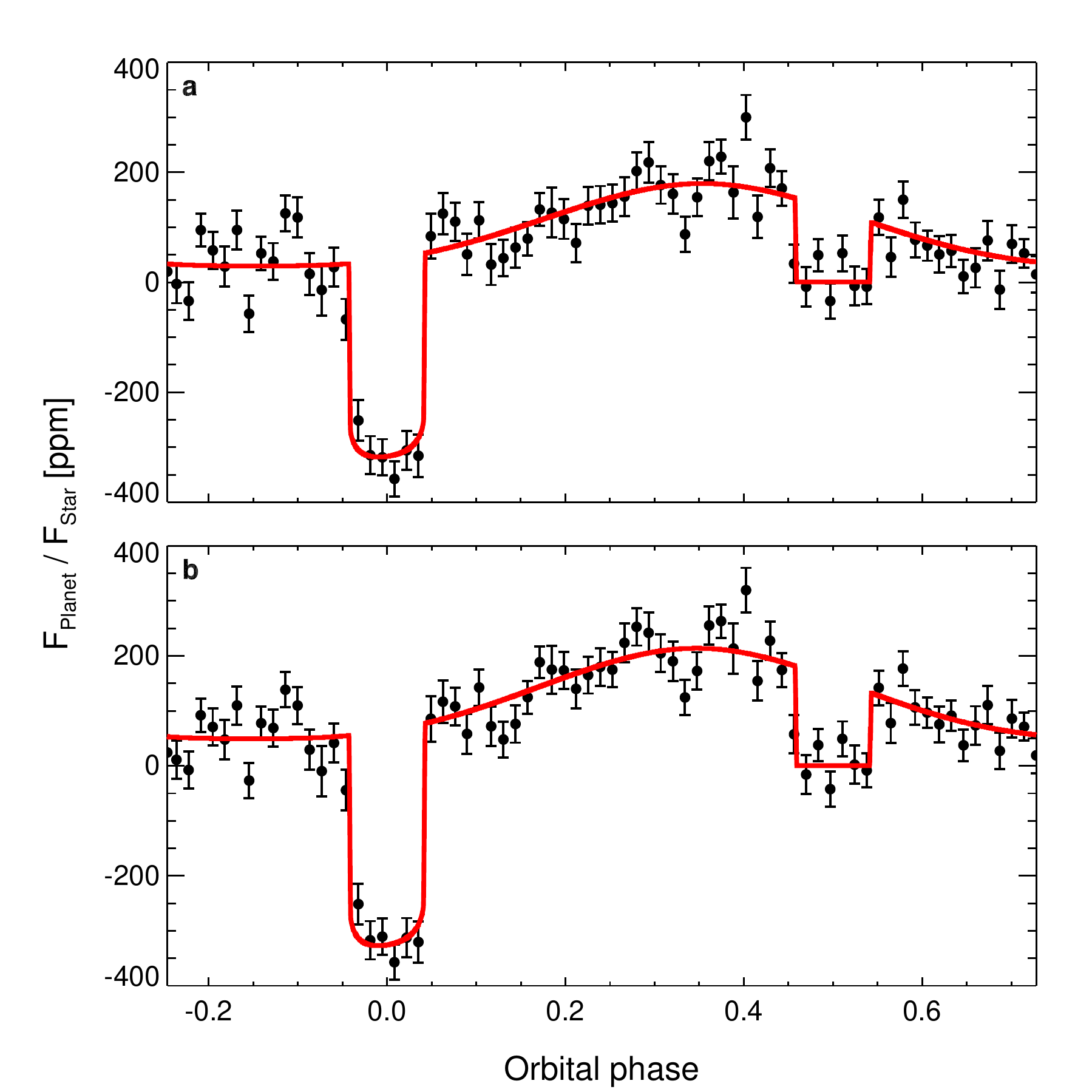}
\caption{\label{fig:poly} | {\bf Polynomial-detrended phase-folded photometry.} Photometry for all 8 datasets combined and folded on 55\,Cancri\,e's orbital period. Contrary to Figure~1, the fits on that figure are obtained using polynomial functions of the centroid position and the PRF's FWHM. Panel (a) shows the fit results employing the entire time-series as input data. Panel (b) shows the results obtained by splitting the times-series in two. Data are binned per 15 minutes. The best-fit model is shown in red on each panel. The error bars are the standard deviation of the mean within each orbital phase bin.}
\end{figure}

% tidal heating
\begin{figure}
\centering
\includegraphics[width=\textwidth]{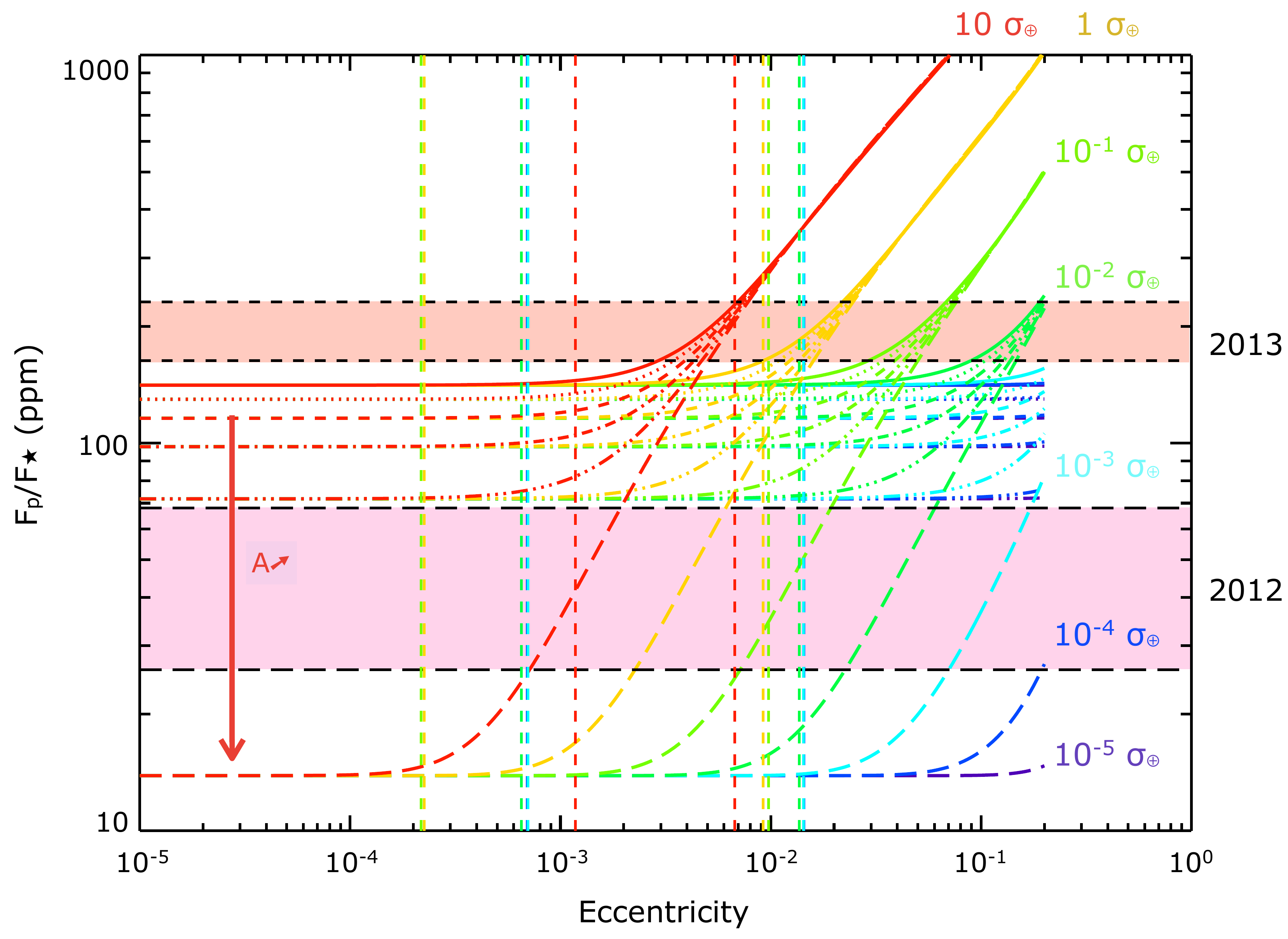}
\caption{\label{fig:tides} | {\bf Tidal heating constraints for 55\,Cancri\,e.} The planet-to-star flux ratio is shown vs. the orbital eccentricity for different values of dissipation (relative to the Earth's, $\sigma_{\oplus}$) and albedos \--- A from 0.0 (solid) to 1.0 (long-dashed). Pink and orange bands represent the occultation depth values measured in 2012 and 2013 with {\it Spitzer}. Vertical lines indicate the plausible range for 55\,Cancri\,e's eccentricity as determined from the n-body simulations for each dissipation value. The 2012 occultation depth can be matched for high albedos and a high dissipation, while the deeper 2013 occultation depth can be matched for the highest dissipation (10$\sigma_{\oplus}$) and the whole albedo range.}
\end{figure}


\begin{thebibliography}{10}

\bibitem{Burrows:2014a}
\bibinfo{author}{{Burrows}, A.~S.}
\newblock \bibinfo{title}{{Highlights in the study of exoplanet atmospheres}}.
\newblock \emph{\bibinfo{journal}{\nat}} \textbf{\bibinfo{volume}{513}},
  \bibinfo{pages}{345--352} (\bibinfo{year}{2014}).

\bibitem{Heng:2015a}
\bibinfo{author}{{Heng}, K.} \& \bibinfo{author}{{Showman}, A.~P.}
\newblock \bibinfo{title}{{Atmospheric Dynamics of Hot Exoplanets}}.
\newblock \emph{\bibinfo{journal}{Annual Review of Earth and Planetary
  Sciences}} \textbf{\bibinfo{volume}{43}}, \bibinfo{pages}{509--540}
  (\bibinfo{year}{2015}).

\bibitem{Knutson:2014a}
\bibinfo{author}{{Knutson}, H.~A.} \emph{et~al.}
\newblock \bibinfo{title}{{Hubble Space Telescope Near-IR Transmission
  Spectroscopy of the Super-Earth HD 97658b}}.
\newblock \emph{\bibinfo{journal}{\apj}} \textbf{\bibinfo{volume}{794}},
  \bibinfo{pages}{155} (\bibinfo{year}{2014}).

\bibitem{Demory:2011}
\bibinfo{author}{Demory, B.-O.} \emph{et~al.}
\newblock \bibinfo{title}{Detection of a transit of the super-earth 55 cancri e
  with warm spitzer}.
\newblock \emph{\bibinfo{journal}{\aap}} \textbf{\bibinfo{volume}{533}},
  \bibinfo{pages}{114} (\bibinfo{year}{2011}).

\bibitem{Winn:2011a}
\bibinfo{author}{{Winn}, J.~N.} \emph{et~al.}
\newblock \bibinfo{title}{{A Super-Earth Transiting a Naked-eye Star}}.
\newblock \emph{\bibinfo{journal}{\apjl}} \textbf{\bibinfo{volume}{737}},
  \bibinfo{pages}{L18} (\bibinfo{year}{2011}).

\bibitem{Solomatov:2007}
\bibinfo{author}{Solomatov, V.}
\newblock \bibinfo{title}{Magma oceans and primordial mantle differentiation}.
\newblock In \bibinfo{editor}{Schubert, G.} (ed.)
  \emph{\bibinfo{booktitle}{Treatise on Geophysics}}, \bibinfo{pages}{91 --
  119} (\bibinfo{publisher}{Elsevier}, \bibinfo{address}{Amsterdam},
  \bibinfo{year}{2007}).

\bibitem{Demory:2015a}
\bibinfo{author}{{Demory}, B.-O.}, \bibinfo{author}{{Gillon}, M.},
  \bibinfo{author}{{Madhusudhan}, N.} \& \bibinfo{author}{{Queloz}, D.}
\newblock \bibinfo{title}{{Variability in the super-Earth 55 Cnc e}}.
\newblock \emph{\bibinfo{journal}{\mnras}} \textbf{\bibinfo{volume}{455}}, \bibinfo{pages}{2018}
  (\bibinfo{year}{2016}).

\bibitem{Gillon:2012a}
\bibinfo{author}{{Gillon}, M.} \emph{et~al.}
\newblock \bibinfo{title}{{The TRAPPIST survey of southern transiting planets.
  I. Thirty eclipses of the ultra-short period planet WASP-43 b}}.
\newblock \emph{\bibinfo{journal}{\aap}} \textbf{\bibinfo{volume}{542}},
  \bibinfo{pages}{A4} (\bibinfo{year}{2012}).

\bibitem{Stevenson:2012a}
\bibinfo{author}{{Stevenson}, K.~B.} \emph{et~al.}
\newblock \bibinfo{title}{{Transit and Eclipse Analyses of the Exoplanet HD
  149026b Using BLISS Mapping}}.
\newblock \emph{\bibinfo{journal}{\apj}} \textbf{\bibinfo{volume}{754}},
  \bibinfo{pages}{136} (\bibinfo{year}{2012}).

\bibitem{Lanotte:2014}
\bibinfo{author}{{Lanotte}, A.~A.} \emph{et~al.}
\newblock \bibinfo{title}{{A global analysis of Spitzer and new HARPS data
  confirms the loneliness and metal-richness of GJ 436 b}}.
\newblock \emph{\bibinfo{journal}{\aap}} \textbf{\bibinfo{volume}{572}},
  \bibinfo{pages}{A73} (\bibinfo{year}{2014}).

\bibitem{Pont:2006b}
\bibinfo{author}{Pont, F.}, \bibinfo{author}{Zucker, S.} \&
  \bibinfo{author}{Queloz, D.}
\newblock \bibinfo{title}{The effect of red noise on planetary transit
  detection}.
\newblock \emph{\bibinfo{journal}{\mnras}} \textbf{\bibinfo{volume}{373}}, \bibinfo{pages}{231}
  (\bibinfo{year}{2006}).

\bibitem{Deming:2015}
\bibinfo{author}{{Deming}, D.} \emph{et~al.}
\newblock \bibinfo{title}{{Spitzer Secondary Eclipses of the Dense,
  Modestly-irradiated, Giant Exoplanet HAT-P-20b Using Pixel-level
  Decorrelation}}.
\newblock \emph{\bibinfo{journal}{\apj}} \textbf{\bibinfo{volume}{805}},
  \bibinfo{pages}{132} (\bibinfo{year}{2015}).

\bibitem{de-Wit:2012a}
\bibinfo{author}{{de Wit}, J.}, \bibinfo{author}{{Gillon}, M.},
  \bibinfo{author}{{Demory}, B.-O.} \& \bibinfo{author}{{Seager}, S.}
\newblock \bibinfo{title}{{Towards consistent mapping of distant worlds:
  secondary-eclipse scanning of the exoplanet HD 189733b}}.
\newblock \emph{\bibinfo{journal}{\aap}} \textbf{\bibinfo{volume}{548}},
  \bibinfo{pages}{A128} (\bibinfo{year}{2012}).

\bibitem{Demory:2013b}
\bibinfo{author}{{Demory}, B.-O.} \emph{et~al.}
\newblock \bibinfo{title}{{Inference of Inhomogeneous Clouds in an Exoplanet
  Atmosphere}}.
\newblock \emph{\bibinfo{journal}{\apjl}} \textbf{\bibinfo{volume}{776}},
  \bibinfo{pages}{L25} (\bibinfo{year}{2013}).

\bibitem{Cowan:2009b}
\bibinfo{author}{{Cowan}, N.~B.} \emph{et~al.}
\newblock \bibinfo{title}{{Alien Maps of an Ocean-bearing World}}.
\newblock \emph{\bibinfo{journal}{\apj}} \textbf{\bibinfo{volume}{700}},
  \bibinfo{pages}{915--923} (\bibinfo{year}{2009}).

\bibitem{Fischer:2008}
\bibinfo{author}{Fischer, D.~A.} \emph{et~al.}
\newblock \bibinfo{title}{Five planets orbiting 55 cancri}.
\newblock \emph{\bibinfo{journal}{\apj}} \textbf{\bibinfo{volume}{675}},
  \bibinfo{pages}{790--801} (\bibinfo{year}{2008}).

\bibitem{Berta:2011}
\bibinfo{author}{{Berta}, Z.~K.} \emph{et~al.}
\newblock \bibinfo{title}{{The GJ1214 Super-Earth System: Stellar Variability,
  New Transits, and a Search for Additional Planets}}.
\newblock \emph{\bibinfo{journal}{\apj}} \textbf{\bibinfo{volume}{736}},
  \bibinfo{pages}{12} (\bibinfo{year}{2011}).

\bibitem{Mazeh:2010a}
\bibinfo{author}{{Mazeh}, T.} \& \bibinfo{author}{{Faigler}, S.}
\newblock \bibinfo{title}{{Detection of the ellipsoidal and the relativistic
  beaming effects in the CoRoT-3 lightcurve}}.
\newblock \emph{\bibinfo{journal}{\aap}} \textbf{\bibinfo{volume}{521}},
  \bibinfo{pages}{L59} (\bibinfo{year}{2010}).

\bibitem{Budaj:2011}
\bibinfo{author}{{Budaj}, J.}
\newblock \bibinfo{title}{{The Reflection Effect in Interacting Binaries or in
  Planet-Star Systems}}.
\newblock \emph{\bibinfo{journal}{\aj}} \textbf{\bibinfo{volume}{141}},
  \bibinfo{pages}{59} (\bibinfo{year}{2011}).

\bibitem{Shkolnik:2008}
\bibinfo{author}{{Shkolnik}, E.}, \bibinfo{author}{{Bohlender}, D.~A.},
  \bibinfo{author}{{Walker}, G.~A.~H.} \& \bibinfo{author}{{Collier Cameron},
  A.}
\newblock \bibinfo{title}{{The On/Off Nature of Star-Planet Interactions}}.
\newblock \emph{\bibinfo{journal}{\apj}} \textbf{\bibinfo{volume}{676}},
  \bibinfo{pages}{628--638} (\bibinfo{year}{2008}).

\bibitem{Miller:2015}
\bibinfo{author}{{Miller}, B.~P.}, \bibinfo{author}{{Gallo}, E.},
  \bibinfo{author}{{Wright}, J.~T.} \& \bibinfo{author}{{Pearson}, E.~G.}
\newblock \bibinfo{title}{{A Comprehensive Statistical Assessment of
  Star-Planet Interaction}}.
\newblock \emph{\bibinfo{journal}{\apj}} \textbf{\bibinfo{volume}{799}},
  \bibinfo{pages}{163} (\bibinfo{year}{2015}).

\bibitem{Showman:2013}
\bibinfo{author}{{Showman}, A.~P.}, \bibinfo{author}{{Fortney}, J.~J.},
  \bibinfo{author}{{Lewis}, N.~K.} \& \bibinfo{author}{{Shabram}, M.}
\newblock \bibinfo{title}{{Doppler Signatures of the Atmospheric Circulation on
  Hot Jupiters}}.
\newblock \emph{\bibinfo{journal}{\apj}} \textbf{\bibinfo{volume}{762}},
  \bibinfo{pages}{24} (\bibinfo{year}{2013}).

\bibitem{Gillon:2012}
\bibinfo{author}{{Gillon}, M.} \emph{et~al.}
\newblock \bibinfo{title}{{Improved precision on the radius of the nearby
  super-Earth 55 Cnc e}}.
\newblock \emph{\bibinfo{journal}{\aap}} \textbf{\bibinfo{volume}{539}},
  \bibinfo{pages}{A28} (\bibinfo{year}{2012}).

\bibitem{Madhusudhan:2010}
\bibinfo{author}{{Madhusudhan}, N.} \& \bibinfo{author}{{Seager}, S.} 
\newblock \bibinfo{title}{{On the Inference of Thermal Inversions in Hot Jupiter Atmospheres}}.
\newblock \emph{\bibinfo{journal}{\apj}} \textbf{\bibinfo{volume}{725}},
  \bibinfo{pages}{261--274} (\bibinfo{year}{2010}).

\bibitem{Ehrenreich:2012}
\bibinfo{author}{{Ehrenreich}, D.} \emph{et~al.}
\newblock \bibinfo{title}{{Hint of a transiting extended atmosphere on 55
  Cancri b}}.
\newblock \emph{\bibinfo{journal}{\aap}} \textbf{\bibinfo{volume}{547}},
  \bibinfo{pages}{A18} (\bibinfo{year}{2012}).

\bibitem{Heng:2012c}
\bibinfo{author}{{Heng}, K.} \& \bibinfo{author}{{Kopparla}, P.}
\newblock \bibinfo{title}{{On the Stability of Super-Earth Atmospheres}}.
\newblock \emph{\bibinfo{journal}{\apj}} \textbf{\bibinfo{volume}{754}},
  \bibinfo{pages}{60} (\bibinfo{year}{2012}).

\bibitem{Schaefer:2011}
\bibinfo{author}{{Schaefer}, L.} \& \bibinfo{author}{{Fegley}, B., Jr.}
\newblock \bibinfo{title}{{Atmospheric Chemistry of Venus-like Exoplanets}}.
\newblock \emph{\bibinfo{journal}{\apj}} \textbf{\bibinfo{volume}{729}},
  \bibinfo{pages}{6} (\bibinfo{year}{2011}).

\bibitem{Miguel:2011}
\bibinfo{author}{Miguel, Y.}, \bibinfo{author}{Kaltenegger, L.},
  \bibinfo{author}{Fegley, B.} \& \bibinfo{author}{Schaefer, L.}
\newblock \bibinfo{title}{Compositions of hot super-earth atmospheres:
  Exploring kepler candidates}.
\newblock \emph{\bibinfo{journal}{\apjl}} \textbf{\bibinfo{volume}{742}},
  \bibinfo{pages}{L19} (\bibinfo{year}{2011}).

\bibitem{Lutgens:2000}
\bibinfo{author}{Lutgens, F.~K.} \& \bibinfo{author}{Tarbuck, E.~J.}
\newblock \emph{\bibinfo{title}{Essentials of Geology}}
  (\bibinfo{publisher}{Prentice Hall,}, \bibinfo{address}{Boston :},
  \bibinfo{year}{2000}), \bibinfo{edition}{7th ed.} edn.

\bibitem{Nelson:2014a}
\bibinfo{author}{{Nelson}, B.~E.} \emph{et~al.}
\newblock \bibinfo{title}{{The 55 Cancri planetary system: fully
  self-consistent N-body constraints and a dynamical analysis}}.
\newblock \emph{\bibinfo{journal}{\mnras}} \textbf{\bibinfo{volume}{441}}, \bibinfo{pages}{442}
  (\bibinfo{year}{2014}).

\setcounter{firstbib}{\value{enumiv}}
\end{thebibliography}

\begin{thebibliography}{10}

\setcounter{enumiv}{\value{firstbib}}

\bibitem{Ballard:2014}
\bibinfo{author}{{Ballard}, S.} \emph{et~al.}
\newblock \bibinfo{title}{{Kepler-93b: A Terrestrial World Measured to within
  120 km, and a Test Case for a New Spitzer Observing Mode}}.
\newblock \emph{\bibinfo{journal}{\apj}} \textbf{\bibinfo{volume}{790}},
  \bibinfo{pages}{12} (\bibinfo{year}{2014}).
  
\bibitem{Eastman:2010a}
\bibinfo{author}{{Eastman}, J.}, \bibinfo{author}{{Siverd}, R.} \&
  \bibinfo{author}{{Gaudi}, B.~S.}
\newblock \bibinfo{title}{{Achieving Better Than 1 Minute Accuracy in the
  Heliocentric and Barycentric Julian Dates}}.
\newblock \emph{\bibinfo{journal}{\pasp}} \textbf{\bibinfo{volume}{122}},
  \bibinfo{pages}{935--946} (\bibinfo{year}{2010}).

\bibitem{Landsman:1993}
\bibinfo{author}{{Landsman}, W.~B.}
\newblock \bibinfo{title}{{The IDL Astronomy User's Library}}.
\newblock In \bibinfo{editor}{{Hanisch}, R.~J.}, \bibinfo{editor}{{Brissenden},
  R.~J.~V.} \& \bibinfo{editor}{{Barnes}, J.} (eds.)
  \emph{\bibinfo{booktitle}{Astronomical Data Analysis Software and Systems
  II}}, vol.~\bibinfo{volume}{52} of \emph{\bibinfo{series}{Astronomical
  Society of the Pacific Conference Series}}, \bibinfo{pages}{246}
  (\bibinfo{year}{1993}).

\bibitem{Agol:2010}
\bibinfo{author}{Agol, E.} \emph{et~al.}
\newblock \bibinfo{title}{The climate of hd 189733b from fourteen transits and
  eclipses measured by spitzer}.
\newblock \emph{\bibinfo{journal}{\apj}} \textbf{\bibinfo{volume}{721}},
  \bibinfo{pages}{1861--1877} (\bibinfo{year}{2010}).

\bibitem{Beerer:2011}
\bibinfo{author}{Beerer, I.~M.} \emph{et~al.}
\newblock \bibinfo{title}{Secondary eclipse photometry of wasp-4b with warm
  spitzer}.
\newblock \emph{\bibinfo{journal}{\apj}} \textbf{\bibinfo{volume}{727}},
  \bibinfo{pages}{23} (\bibinfo{year}{2011}).

\bibitem{Schwarz:1978}
\bibinfo{author}{Schwarz, G.}
\newblock \bibinfo{title}{Estimating the dimension of a model}.
\newblock \emph{\bibinfo{journal}{The Annals of Statistics}}
  \textbf{\bibinfo{volume}{6}}, \bibinfo{pages}{461--464}
  (\bibinfo{year}{1978}).

\bibitem{Sobolev:1975}
\bibinfo{author}{Sobolev, V.V.}
\newblock \bibinfo{title}{Light scattering in planetary atmospheres}.
\newblock \emph{\bibinfo{book}{(Translation of Rasseianie sveta v atmosferakh planet, Moscow, Izdatel'stvo Nauka, 1972.) Oxford and New York, Pergamon Press (International Series of Monographs in Natural Philosophy)}}
  \textbf{\bibinfo{volume}{76}}, \bibinfo{pages}{461--464}
  (\bibinfo{year}{1975}).

\bibitem{Mandel:2002}
\bibinfo{author}{{Mandel}, K.} \& \bibinfo{author}{{Agol}, E.}
\newblock \bibinfo{title}{Analytic Light Curves for Planetary Transit Searches}.
\newblock \emph{\bibinfo{journal}{\apj}} \textbf{\bibinfo{volume}{580}},
  \bibinfo{pages}{L171} (\bibinfo{year}{2002}).

\bibitem{Claret:2011}
\bibinfo{author}{{Claret}, A.} \& \bibinfo{author}{{Bloemen}, S.}
\newblock \bibinfo{title}{{Gravity and limb-darkening coefficients for the Kepler, CoRoT, Spitzer, uvby, UBVRIJHK, and Sloan photometric systems}}.
\newblock \emph{\bibinfo{journal}{\aap}} \textbf{\bibinfo{volume}{529}},
  \bibinfo{pages}{75} (\bibinfo{year}{2011}).

\bibitem{von-Braun:2011a}
\bibinfo{author}{{von Braun}, K.} \emph{et~al.}
\newblock \bibinfo{title}{55 Cancri: Stellar Astrophysical Parameters, a Planet in the Habitable Zone, and Implications for the Radius of a Transiting Super-Earth}.
\newblock \emph{\bibinfo{journal}{\apj}} \textbf{\bibinfo{volume}{740}},
  \bibinfo{pages}{49} (\bibinfo{year}{2011}).

\bibitem{Knutson:2007}
\bibinfo{author}{Knutson, H.~A.} \emph{et~al.}
\newblock \bibinfo{title}{A map of the day-night contrast of the extrasolar
  planet hd 189733b}.
\newblock \emph{\bibinfo{journal}{Nature}} \textbf{\bibinfo{volume}{447}},
  \bibinfo{pages}{183} (\bibinfo{year}{2007}).

\bibitem{Cowan:2008}
\bibinfo{author}{{Cowan}, N.~B.} \& \bibinfo{author}{{Agol}, E.}
\newblock \bibinfo{title}{{Inverting Phase Functions to Map Exoplanets}}.
\newblock \emph{\bibinfo{journal}{\apjl}} \textbf{\bibinfo{volume}{678}},
  \bibinfo{pages}{L129--L132} (\bibinfo{year}{2008}).

\bibitem{Crossfield:2012d}
\bibinfo{author}{{Crossfield}, I.~J.~M.}
\newblock \bibinfo{title}{{ACME stellar spectra. I. Absolutely calibrated,
  mostly empirical flux densities of 55 Cancri and its transiting planet 55
  Cancri e}}.
\newblock \emph{\bibinfo{journal}{\aap}} \textbf{\bibinfo{volume}{545}},
  \bibinfo{pages}{A97} (\bibinfo{year}{2012}).

\bibitem{Menou:2012b}
\bibinfo{author}{{Menou}, K.}
\newblock \bibinfo{title}{{Magnetic Scaling Laws for the Atmospheres of Hot
  Giant Exoplanets}}.
\newblock \emph{\bibinfo{journal}{\apj}} \textbf{\bibinfo{volume}{745}},
  \bibinfo{pages}{138} (\bibinfo{year}{2012}).

\bibitem{Owen:2013}
\bibinfo{author}{{Owen}, J.~E.} \& \bibinfo{author}{{Wu}, Y.}
\newblock \bibinfo{title}{{Kepler Planets: A Tale of Evaporation}}.
\newblock \emph{\bibinfo{journal}{\apj}} \textbf{\bibinfo{volume}{775}},
  \bibinfo{pages}{105} (\bibinfo{year}{2013}).

\bibitem{Bolmont:2015}
\bibinfo{author}{{Bolmont}, E.}, \bibinfo{author}{{Raymond}, S.~N.},
  \bibinfo{author}{{Leconte}, J.}, \bibinfo{author}{{Hersant}, F.} \&
  \bibinfo{author}{{Correia}, A.~C.~M.}
\newblock \bibinfo{title}{{Mercury-T: A new code to study tidally evolving
  multi-planet systems. Applications to Kepler-62}}.
\newblock \emph{\bibinfo{journal}{ArXiv e-prints}}  (\bibinfo{year}{2015}).

\end{thebibliography}
\end{document}